\documentclass[acmsmall]{acmart}

\AtBeginDocument{%
  \providecommand\BibTeX{{%
    \normalfont B\kern-0.5em{\scshape i\kern-0.25em b}\kern-0.8em\TeX}}}

\usepackage{multirow}
\usepackage{makecell}
\usepackage{mathrsfs,amsmath}

\begin{document}

\title{Network Representation Learning: From Preprocessing, Feature Extraction to Node Embedding}

\author{JINGYA ZHOU}
\email{jy\_zhou@suda.edu.cn}
\orcid{0000-0003-0721-7424}
\authornotemark[1]
\affiliation{%
  \institution{Soochow University}
  \streetaddress{1 Shizi street}
  \city{Suzhou}
  \state{Jiangsu}
  \postcode{215006}
  \country{China}
  \institution{Georgia Institute of Technology}
  \streetaddress{during May 2019 - June 2020}
  \city{Atlanta}
  \state{Georgia}
  \country{USA}
  \institution{State Key Laboratory of Mathematical Engineering and Advanced Computing}
  \city{Wuxi}
  \state{Jiangsu}
  \postcode{214125}
  \country{China}
}

\author{LING LIU}
\affiliation{%
  \institution{Georgia Institute of Technology}
  \streetaddress{801 Atlantic Drive}
  \city{Atlanta}
  \state{Georgia}
  \postcode{30332}
  \country{USA}
}
\email{lingliu@cc.gatech.edu}

\author{WENQI WEI}
\affiliation{%
	\institution{Georgia Institute of Technology}
	\streetaddress{801 Atlantic Drive}
	\city{Atlanta}
	\state{Georgia}
	\postcode{30332}
	\country{USA}
}
\email{wenqiwei@gatech.edu}

\author{JIANXI FAN}
\affiliation{%
	\institution{Soochow University}
	\streetaddress{1 Shizi street}
	\city{Suzhou}
	\state{Jiangsu}
	\country{China}
	\postcode{215006}}
\email{jxfan@suda.edu.cn}

\renewcommand{\shortauthors}{J. Zhou et al.}

\newcommand{\tabincell}[2]{\begin{tabular}{@{}#1@{}}#2\end{tabular}} 
\begin{abstract}
	Network representation learning (NRL) advances the conventional graph mining of social networks, knowledge graphs, and complex biomedical and physics information networks. Over dozens of network representation learning algorithms have been reported in the literature. Most of them focus on learning node embeddings for homogeneous networks, but they differ in the specific encoding schemes and specific types of node semantics captured and used for learning node embedding. This survey paper reviews the design principles and the different node embedding techniques for network representation learning over homogeneous networks. To facilitate the comparison of different node embedding algorithms, we introduce a unified reference framework to divide and generalize the node embedding learning process on a given network into preprocessing steps, node feature extraction steps and node embedding model training for a NRL task such as link prediction and node clustering. With this unifying reference framework, we highlight the representative methods, models, and techniques used at different stages of the node embedding model learning process. This survey not only helps researchers and practitioners to gain an in-depth understanding of different network representation learning techniques but also provides practical guidelines for designing and developing the next generation of network representation learning algorithms and systems.
\end{abstract}

\ccsdesc[500]{Computing methodologies}
\ccsdesc[500]{Computing methodologies~Machine learning}
\ccsdesc[500]{Information systems~Data mining}

\keywords{Network representation learning, data preprocessing, feature extraction, node embedding}

\maketitle

\section{Introduction}
Recent advances in deep learning and convolutional neural network (CNN) \cite{DBLP:conf/nips/KrizhevskySH12} have made remarkable breakthroughs to many fields, such as machine translation \cite{DBLP:journals/corr/WuSCLNMKCGMKSJL16} and reading comprehension in  natural language processing (NLP) \cite{DBLP:conf/iclr/YuDLZ00L18},  object detection \cite{DBLP:conf/nips/SzegedyTE13} and image classification \cite{DBLP:conf/cvpr/MarinoSG17} in computer vision (CV). In addition to text, audio, image and video data, information networks (or graphs) represent another type of natural and complex data structure representing a set of entities and their relationships. A wide variety of real-world data in business, science and engineering domains are best captured as information networks, such as protein interaction networks, citation networks, and social media networks like Facebook, LinkedIn, to name a few. 

Network representation learning (NRL), also known as network embedding, is to train a neural network to represent an information network as a collection of node-embedding vectors in a latent space such that the desired network features are preserved, which enables the well-trained NRL model to perform network analytics, such as link prediction or node cluster, as shown in Fig. 1. The goal of NRL is to employ deep learning algorithms to encode useful network information into the latent semantic representations, which can be deployed for performing popular network analytics, such as node classification, link prediction, community detection, and domain-specific network mining, such as social recommendation \cite{DBLP:conf/www/Fan0LHZTY19,DBLP:conf/ijcai/WangLDWZ19}, protein to protein interaction prediction \cite{DBLP:conf/nips/FoutBSB17}, disease-gene association identification \cite{DBLP:conf/kdd/HanYZSLZ0K19}, automatic molecule optimization \cite{DBLP:conf/aaai/FuXS20} and Bitcoin transaction forecasting \cite{DBLP:journals/tdsc/Weiwq20}.
\begin{figure}[t]
	\centering
	\includegraphics[width=\linewidth]{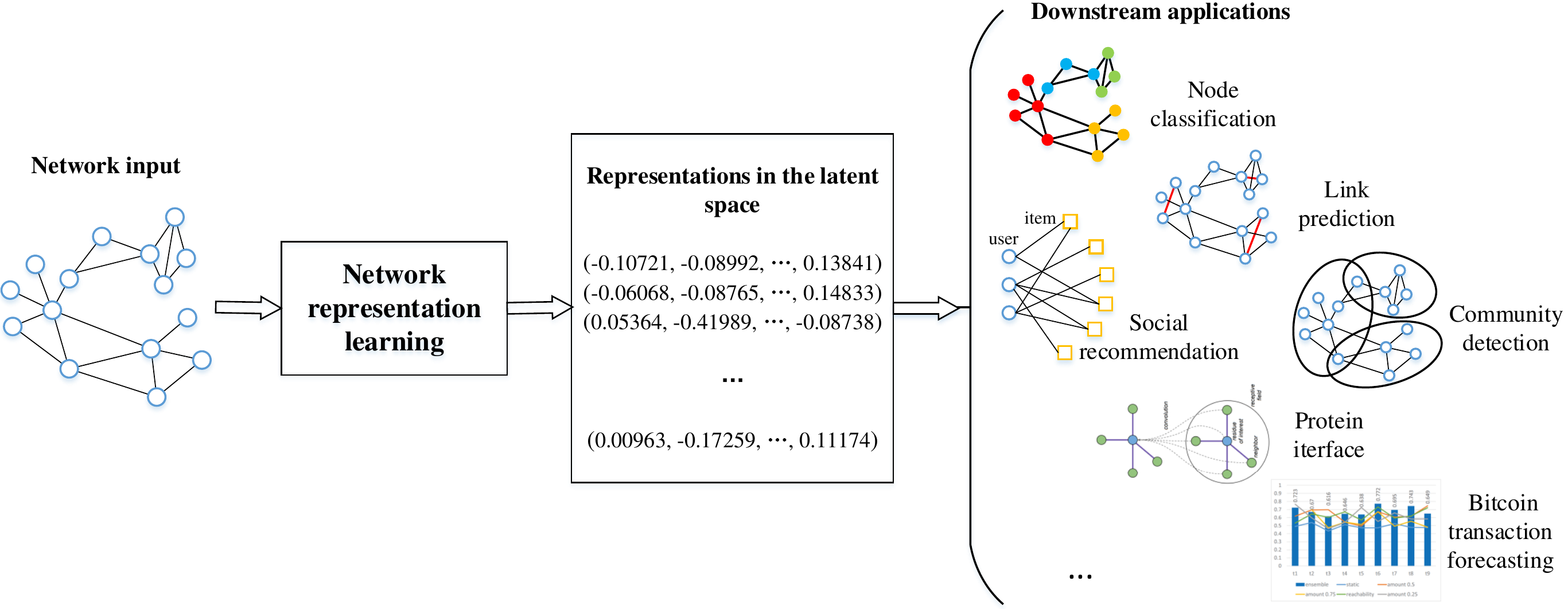}
	\caption{An illustration of NRL and its deployment in example applications \cite{DBLP:conf/www/Fan0LHZTY19,DBLP:conf/ijcai/WangLDWZ19,DBLP:conf/nips/FoutBSB17,DBLP:conf/kdd/HanYZSLZ0K19,DBLP:conf/aaai/FuXS20,DBLP:journals/tdsc/Weiwq20}.}
	\Description{NRL introduction}
\end{figure}

Different from traditional feature engineering that relies heavily on handcrafted statistics to extract structural information, NRL introduces a new data-driven deep learning paradigm to capture, encode and embed structural features along with non-structural features into a latent space represented by dense and continuous vectors. By embedding edge semantics into node vectors, a variety of network operations can be carried out efficiently, e.g., computing the similarity between a pair of nodes, visualizing a network in a 2-dimensional space. Moreover, parallel processing on large scale networks can be naturally supported with the node embedding learned from NRL. 

Most of the existing network representation learning efforts are targeted on learning node embeddings of a homogeneous network, in which all nodes are homogeneous and all edges belong to a single type of node relationships, e.g., a social network is considered homogeneous when we only consider users and their friendship relationships \cite{DBLP:conf/kdd/PerozziAS14}. A heterogeneous information network consists of nodes and edges of heterogeneous types, corresponding to different types of entities and different kinds of relations respectively. Knowledge graph~\cite{DBLP:conf/aaai/GuoWWWG18,DBLP:journals/corr/abs-2002-00388} and RDF graphs~\cite{DBLP:journals/pvldb/YuanLWJZL13} are known examples of heterogeneous information networks.

DeepWalk~\cite{DBLP:conf/kdd/PerozziAS14} is the first node embedding algorithm that learns to encode the neighborhood features of each node in a homogeneous graph through learning the encoding of its scoped random walk properties using the autoencoder algorithms in conjunction with node2vec~\cite{DBLP:conf/kdd/GroverL16}. Inspired by DeepWalk design, dozens of node embedding algorithms have been proposed~\cite{DBLP:conf/kdd/GroverL16,DBLP:conf/kdd/RibeiroSF17,DBLP:conf/kdd/DonnatZHL18,DBLP:conf/asunam/RozemberczkiDSS19,DBLP:conf/icdm/ZhangYZZ18,DBLP:conf/www/EpastoP19,DBLP:conf/kdd/LiuTLYZH19,DBLP:conf/www/TangQWZYM15,DBLP:conf/aaai/WangWWZZZXG18,DBLP:conf/cikm/CaoLX15,DBLP:conf/ijcai/YangLZSC15,DBLP:conf/kdd/WangC016,DBLP:conf/aaai/CaoLX16,DBLP:conf/cikm/QuTSR0017,DBLP:conf/www/TsitsulinMKM18,DBLP:conf/nips/HamiltonYL17,DBLP:conf/iclr/BojchevskiG18,DBLP:conf/ijcai/GuoXL19,DBLP:conf/www/RossiZA18,DBLP:conf/ijcai/SunWHTH19,DBLP:iclr/abs-1910-02370,DBLP:conf/iclr/KipfW17, DBLP:conf/kdd/WuHX19, DBLP:conf/nips/ZhangC18,DBLP:conf/kdd/CenZZYZ019,DBLP:journals/tkde/LiaoHZC18,DBLP:conf/asunam/PerozziKCS17}. Although most of them focus on learning node embeddings for homogeneous networks, they differ in terms of the specific encoding schemes and the specific types of node semantics captured and used for learning node embedding. This survey paper mainly reviews the design principles and the different node embedding techniques developed for network representation learning over homogeneous networks. To facilitate the comparison of different node embedding algorithms, we introduce a unified reference framework to divide and generalize the node embedding learning process on a given network into preprocessing steps, node feature extraction steps and node embedding model that can be used for link prediction and node clustering. With this unifying reference framework, we highlight the most representative methods, models, and techniques used at different stages of the node embedding model learning process.

We argue that an in-depth understanding of different node embedding methods/models/techniques is also essential for other types of network representation learning approaches that are built on top of node embedding techniques, such as edge embedding \cite{DBLP:conf/cikm/Abu-El-HaijaPA17,DBLP:journals/bmcbi/GaoFOTLYGFWDY19}, subgraph embedding \cite{DBLP:conf/emnlp/BordesCW14,DBLP:conf/aaai/CaoWM18} and entire-graph embedding \cite{DBLP:conf/ijcai/BaiDQMG0SW19,DBLP:journals/corr/NarayananCVCLJ17}. For example, an edge can be represented by a Hadamard product of its two adjacent nodes' vectors. Similarly, graph coarsening mechanisms \cite{DBLP:conf/aaai/ChenPHS18,DBLP:conf/cogmi/YuZDLPCGTMIS19} may create a hierarchy by successively clustering the nodes in the input graph into smaller graphs connected in a hierarchical manner, which can be used to generate representations for subgraphs and even for the entire graph.

We conjecture that this survey paper not only helps researchers and practitioners to gain an in-depth understanding of different network representation learning techniques, but also provides practical guidelines for designing and developing the next generation of network representation learning algorithms and systems. 

Current surveys \cite{DBLP:journals/tkde/CaiZC18,DBLP:journals/debu/HamiltonYL17,DBLP:journals/tkde/CuiWPZ19} primarily focus on presenting a taxonomy to review the existing work on network representation learning. 
	Concretely, \cite{DBLP:journals/tkde/CaiZC18} proposes two taxonomies of graph embedding based on problem settings and techniques respectively and it first appeared in 2017 on ArXiv and published in 2018. \cite{DBLP:journals/tkde/CuiWPZ19} proposes a taxonomy of network embedding according to the types of information preserved. \cite{DBLP:journals/debu/HamiltonYL17} appeared in 2017 in the IEEE Data Eng. Bulletin. It describes a set of conventional node embedding methods with the focus on pairwise proximity methods and neighborhood aggregation based methods.
	In contrast, our unified reference framework provides a broader and more comprehensive comparative review of the state of the art in network representation learning. In our three-stage reference framework, each stage serves as a categorization of the set of technical solutions dedicated to the tasks respective to this stage. For example, we not only provide a review of node embedding models using the unified framework, but also describe a set of optimization techniques that are commonly used in different node embedding methods, and also an overview of recent advances in NRL.	

We list the mathematical notations used throughout the paper in Table 1 for reference convenience.

The remainder of this survey is structured as follows. In Section 2, we describe the basic steps of network representation learning to generate node embeddings using the autoencoder approach. In Section 3, we present an overview of the unifying three-stage reference framework for NRL, and discuss representative methods, models, and optimization techniques used at each stage. In Section 4, we review recent advances in conventional NRL, distributed NRL, multi-NRL, dynamic NRL and knowledge graph representation learning by using the proposed reference framework and discuss several open challenges. In Section 5, we conclude our survey.

\begin{table}[t]
	\centering
	\tiny
	\setlength{\abovecaptionskip}{2pt}
	\caption{The list of notations and symbols used in the paper and their meaning}
	\begin{tabular}{|l|l|l|l|}
		\hline
		\textbf{Notation/Symbol} & \textbf{Meaning} & \textbf{Notation/Symbol} & \textbf{Meaning} \\
		\hline
		\textbf{W, U, M} & bold capital letters represent matrices & \textbf{h}, ${\bf{u}}_k$, ${\bf{v}_I}$  & bold lowercase letters represent vectors \\
		${\rm{u}}_k$, ${\rm{v}_I}$, ${\rm{v}}_i$, ${\bf{v}}_j$ & lowercase letters represent nodes & $v_i$, $h_i$, $y_i$ & italic lowercase letters represent vector elements \\
		$\Phi ({\cdot}{\rm{)}}$, $f_{\theta_1}(\cdot)$ & mapping function, i.e., encoder &  $\Psi (\cdot)$, $g_{\theta_2}(\cdot)$  &  decoder \\
		$\bf A$ & adjacency matrix & $\bf L$ & Laplacian matrix \\
		$N_{\rm u}$ & node $\rm u$'s neighborhood & $G_p$ & persona graph \\
		$G[N_{\rm u}]$ & ego-network of node $\rm u$ & $\bf P^k$, $\bf{\Gamma}^k$ & $k$-step transition matrix \\
		$\bf X$ & additional information matrix & ${{\hat \Psi }}({{\bf{v}}_{\bf{i}}},{{\bf{v}}_{\bf{j}}})$ & empirical probability \\
		$\bf{g}_\theta$ & filter of a convolution operation & $T_k(\bf x)$ & Chebyshev polynomials \\
		${\bf{\Theta }}$ & matrix of filter parameters & $AGG_k$ & aggregation function in the $k$th layer \\
		${\rm V}_{neg}$ & set of negative samples & $\eta$ & learning rate \\
		$L_{1st}$, $L_{2nd}$, $L$ & first loss, second loss, loss & $\bf b$ & biase vector \\
		${\bf x}^{(i)}$,  ${\bf{\tilde x}}^{(i)}$ & $i$th instance and its corrupted form & $\bf(H^{(i)})_l$ & matrix of representations in the $l$th layer of network $i$ \\
		$\bf S^{RPR}$ & rooted PageRank matrix & $\bar t_p$ & timestamp before the current event \\
		$\bf h$, $\bf r$, $\bf t$ & representations of head, translation and tail & $d_r(\bf h, t)$ & distance function of a triple \\
		\hline
	\end{tabular}%
	\label{tab1}%
\end{table}%

\section{Network Representation Learning: What and How}
To establish a common ground for introducing design principles of NRL methods and techniques, we first provide a walkthrough example to illustrate how network representation learning works from the autoencoder perspective. We first briefly describe DeepWalk \cite{DBLP:conf/kdd/PerozziAS14} as it will be used as the reference NRL model in this section. 

DeepWalk~\cite{DBLP:conf/kdd/PerozziAS14} is generalized from the advancements in language modeling, e.g., word2vec \cite{DBLP:conf/nips/MikolovSCCD13}. In a language model, the corpus is built by collecting sentences from many documents. If we regard node traveling paths as sentences, we can build corpus for network representation learning. Given an input network, we use the random walk method with parameters $\gamma$, $t$ and $\alpha$ to generate multiple node traveling paths, where $\gamma$ decides how many times to issue random walks from a node, $t$ is the path length, and $\alpha$ is the probability of stopping walk and restarting from the initial node.
\begin{figure}[t]
	\centering
	\includegraphics[width=\linewidth]{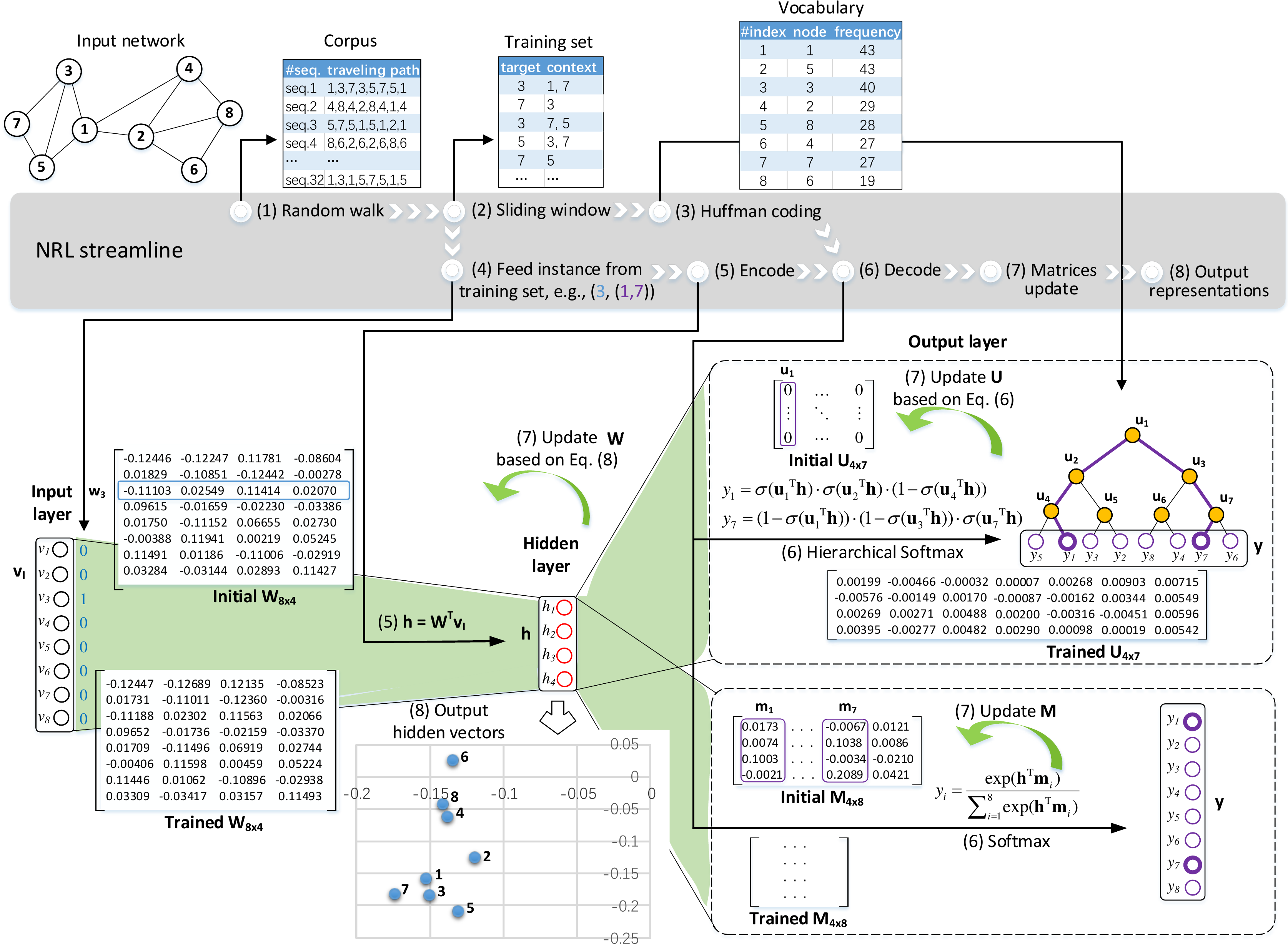}
	\caption{An example of a three-layer neural network used to realize network representation learning. (1) build corpus based on random walk ($\gamma =4$, $t=8$, $\alpha=0$); (2) build training set based on sliding window ($s=1$); (3) build vocabulary to store the set of nodes and their frequency in corpus, and build a Huffman tree. Note that this step is optional and only necessary for hierarchical softmax; (4) feed training instance to input layer; (5) encode input vector into hidden vector; (6) decode hidden vector into conditional probabilities of context nodes based on softmax or hierarchical softmax; (7) update matrices \textbf{W} and \textbf{M} or \textbf{U} based on back-propagation method ($\eta =0.025$); (8) output hidden vectors as the learned representations.}
	\Description{NRL example}
\end{figure}

In the language processing field, skip-gram and continuous bag-of-word (CBOW) \cite{DBLP:conf/nips/MikolovSCCD13} are two commonly used models for estimating the likelihood of co-occurrence among words in training set. For network representation learning, training instances are extracted from node traveling paths based on a sliding window. An instance consists of a target node and its context located within a fixed window size, e.g., (3, (1,7)). In the meantime, we also need to build a vocabulary to index and sort all nodes by their frequency in the corpus, and then build a Huffman tree based on the frequency for hierarchical softmax.

The learning model shown in Fig. 2 contains three layers and it belongs to a typical autoencoder paradigm. The input vectors are encoded into latent representation vectors by means of a matrix \textbf{W}, which is known as encoding, and then the latent representation vectors are reconstructed into output vectors by means of a matrix \textbf{U}, which is known as decoding. Given a training instance, skip-gram model is mainly used to forecast the context given a target node, while on the contrary CBOW model predicts the target node given its context. Skip-gram model is widely adopted in network representation learning, since the conditional probability can be decomposed into multiple simple conditional probabilities under independence assumption,
\begin{equation}\label{1}
\Pr (context({{\rm{v}}_{\rm{I}}})|\Phi ({{\rm{v}}_{\rm{I}}}{\rm{)}}) = \prod\limits_{{{\rm{v}}_j} \in contex({{\rm{v}}_{\rm{I}}})} {\Pr ({{\rm{v}}_j}|\Phi ({{\rm{v}}_{\rm{I}}}{\rm{)}})} ,
\end{equation}
where $\Phi ({\cdot}{\rm{)}}$ is a mapping function that embeds nodes into the low-dimensional vector space, i.e., $\Phi ({{\rm{v}}_{\rm{I}}}{\rm{)}}$ refers to the target node $\rm{v_I}$'s learned representation. Here $\Phi ({\cdot}{\rm{)}}$ acts as the encoder from the autoencoder perspective. 
Given the embedding of the target node, the conditional probability acts as the decoder that captures the reconstruction of the target node and its context nodes in the original network. A loss function is defined accordingly to measure the reconstruction error and the learning model is trained by minimizing the loss. To be specific, for each training instance, the target node $\rm{v_I}$ is initially encoded to be a one-hot vector $\bf{v_I}$ and its dimension equals vocabulary size $n$, and \textbf{W} is a $n \times m$ matrix initialized by letting its entries randomly falling in a range $[-1/2n, 1/2n]$. One-hot encoding implies that $\bf{h}$ is a $m$-dimensional vector simply copying a row of \textbf{W} associated with the target node $\rm{v_I}$. A $m \times n$ matrix $\bf{M}$ is set to decode the encoded vector $\bf{v_I}$, where the conditional probability is obtained by doing softmax, i.e., ${y_i} = \frac{{\exp ({{\bf{h}}^{\rm{T}}}{{\bf{m}}_i})}}{{\sum\nolimits_{i = 1}^n {\exp ({{\bf{h}}^{\rm{T}}}{{\bf{m}}_i})} }}$. But softmax is not scalable, because for each training instance, softmax requires to repeat vector multiplication for $n$ times to obtain the denominator.

To improve computation efficiency of decoder, DeepWalk uses hierarchical softmax \cite{DBLP:journals/corr/abs-1301-3781,DBLP:conf/aistats/MorinB05} instead of softmax to implement the conditional probability factorization. Hierarchical softmax model builds a binary tree and places all network nodes on the leaf layer. Then there will be $n-1$ branch nodes and each of them has an associated $m$-dimensional vector. For the output node $\rm{v}_j$ in a training instance, it corresponds to a leaf node $y_j$ in the tree representing the probability of $\rm{v}_j$ in the output layer given target node $\rm{v_I}$. It is easy to identify a unique path from the root node to node $y_i$, and the conditional probability can be computed based on the path, i.e.,
\begin{equation}\label{2}
{y_j} = \Pr ({{\rm{v}}_j}|\Phi ({{\rm{v}}_{\rm{I}}}{\rm{)}}) = \prod\limits_{k = 1}^{{l_j} - 1} {\Pr ({{\bf{u}}_k}|\Phi ({{\rm{v}}_{\rm{I}}}{\rm{)}})} ,
\end{equation}
where $l_j$ is the length of the path toward $y_j$. DeepWalk uses Huffman tree to implement hierarchical softmax due to its optimal property on average path length. On the path toward leaf node $y_j$, a binary classifier is used to compute the probability of going left or right at each branch node, i.e.,
\begin{equation}\label{3}
\Pr ({{\bf{u}}_k}|\Phi ({{\rm{v}}_{\rm{I}}}{\rm{)}}) = \left\{ \begin{array}{l}
\sigma ({{\bf{u}}_k}^{\rm{T}} \cdot {\bf{h}}),{\rm{go \, left}} \\ 
{\textcolor{black}{1-}}\sigma ({{\bf{u}}_k}^{\rm{T}} \cdot {\bf{h}}),{\rm{go \, right}} \\ 
\end{array} \right.,{\, \rm{where }}\, \sigma ({{\bf{u}}_k}^{\rm{T}} \cdot {\bf{h}}) = \frac{1}{{1 + \exp ( - {{\bf{u}}_k}^{\rm{T}} \cdot {\bf{h}})}}.
\end{equation}

The model's goal is to obtain the maximized conditional probability, which is equivalent to minimize the following loss function
\begin{equation}\label{4}
L =  - \log \Pr (context({{\rm{v}}_{\rm{I}}})|\Phi ({{\rm{v}}_{\rm{I}}}{\rm{)}}) = \sum\limits_{{{\rm{v}}_j} \in context({{\rm{v}}_{\rm{I}}})} { - \log {y_j}}.
\end{equation}
To this end, back-propagation method is used to update two weight matrices \textbf{W} and \textbf{U} with gradient descent. Firstly, we take the derivative of loss with regard to each $\bf{u_k}$ on the path toward the context node and obtain
\begin{equation}\label{5}
\frac{{\partial L}}{{\partial {{\bf{u}}_k}}} = \frac{{\partial L}}{{\partial {{\bf{u}}_k}^{\rm{T}} \cdot {\bf{h}}}} \cdot \frac{{\partial {{\bf{u}}_k}^{\rm{T}} \cdot {\bf{h}}}}{{\partial {{\bf{u}}_k}}} = \sum\limits_{{{\rm{v}}_j} \in context({{\rm{v}}_{\rm{I}}})} {\left( {\sigma ({{\bf{u}}_k}^{\rm{T}} \cdot {\bf{h}}) - {\pi _k}} \right)}  \cdot {\bf{h}},
\end{equation}
where $\pi_k =1$ if go left and $\pi_k=0$ otherwise. The corresponding vectors in matrix \textbf{U} is updated by
\begin{equation}\label{6}
{{\bf{u}}_k} \leftarrow {{\bf{u}}_k} - \eta \sum\limits_{{{\rm{v}}_j} \in context({{\rm{v}}_{\rm{I}}})} {\left( {\sigma ({{\bf{u}}_k}^{\rm{T}} \cdot {\bf{h}}) - {\pi _k}} \right)}  \cdot {\bf{h}}, \, {\rm{where}} \, \eta \, {\rm{ is \, the \, learning \, rate}}{\rm{.}}
\end{equation}
Then for each context node in an instance, we take the derivative of loss with regard to the hidden vector $\bf{h}$ and obtain
\begin{equation}\label{7}
\frac{{\partial L}}{{\partial {\bf{h}}}} = \sum\limits_{{{\rm{v}}_j} \in context({{\rm{v}}_{\rm{I}}})} {\sum\limits_{k = 1}^{{l_j} - 1} {\frac{{\partial L}}{{\partial {{\bf{u}}_k}^{\rm{T}} \cdot {\bf{h}}}} \cdot \frac{{\partial {{\bf{u}}_k}^{\rm{T}} \cdot {\bf{h}}}}{{\partial {\bf{h}}}}} }  = \sum\limits_{{{\rm{v}}_j} \in context({{\rm{v}}_{\rm{I}}})} {\sum\limits_{k = 1}^{{l_j} - 1} {\left( {\sigma ({{\bf{u}}_k}^{\rm{T}} \cdot {\bf{h}}) - {\pi _k}} \right) \cdot } {{\bf{u}}_k}} .
\end{equation}
The vector $\bf{w_I}$ in matrix \textbf{W} is updated accordingly by
\begin{equation}\label{8}
{{\bf{w}}_{\rm{I}}} \leftarrow {{\bf{w}}_{\rm{I}}} - \eta \sum\limits_{{{\rm{v}}_j} \in context({{\rm{v}}_{\rm{I}}})} {\sum\limits_{k = 1}^{{l_j} - 1} {\left( {\sigma ({{\bf{u}}_k}^{\rm{T}} \cdot {\bf{h}}) - {\pi _k}} \right) \cdot } {{\bf{u}}_k}} .
\end{equation}
The model will learn the latent representation for every node by updating matrices iteratively and eventually stabilize. The hidden vectors are the node representations learned from the network.

Fig. 2 shows a simple network with 8 nodes, we build a corpus by running random walk 4 times for each node and obtain 32 node sequences. We also generate multiple instances for training by means of a sliding window with $s=1$, which means a target node may have 2 context nodes at most. Instance (3, (1,7)) is the first one used for the following training, and the target node's input vector is $\rm{v_3}$'s one-hot vector (0,0,1,0,0,0,0,0). After encoded by weight matrix \textbf{W}, we obtain $\rm{v_3}$'s hidden vector that is $\bf{w_3}$ here. In order to obtain the conditional probability, we build a Huffman tree in the output layer. The weight matrix \textbf{U} is a collection of all branch nodes' vectors, and they are initialized to be zero vectors so as to make sure that the probabilities of going left and going right at each branch are initially identical, i.e., $1/2$. Paths ($\bf{u_1}$, $\bf{u_2}$, $\bf{u_4}$) and ($\bf{u_1}$, $\bf{u_3}$, $\bf{u_7}$) are two unique paths toward leaf nodes $y_1$ and $y_7$ respectively, and then we have $y_1=y_7=1/8$ based on Eq. (2). In the following process, we need to update the correlated vectors in weight matrices reversely to minimize the loss, so we obtain $\bf{u_1}=\bf{0}$, $\bf{u_2}=\bf{u_7}=\eta \bf{h}/2$, $\bf{u_4}=\bf{u_3}=-\eta \bf{h}/2$ based on Eq. (6), and then $\bf{w_3}=\bf{w_3}^{(old)} - \eta^2\bf{h}/2$ based on Eq. (8). After we finish the training against all instances, the representation learned from the model is the hidden vector. Since the input layer uses one-hot encoding, the network representation is weight matrix \textbf{W}. We plot these representations in a two-dimensional coordinate system for visualization, where community structure can be observed.


\section{The Reference Framework for Network Representation Learning}
We present a unified reference framework illustrated in Fig. 3 to capture the workflow of network representation learning and it contains three consecutive stages.
The first stage is the network data preprocessing, and it is responsible for obtaining the desired network structure information from the original input network. During this stage, the prime aim is to employ a learning-task suitable network preprocessing method to transform the input network into a set of internal data structures, which is more suitable for structural feature extraction in the next stage. Node state and edge state provide additional context information other than network topological structure, which are useful and can be leveraged for learning network representations. Different NRL algorithms tend to have different design choices on which additional information will be utilized to augment the node context information in addition to network topological structure.
The second stage is the network feature extraction and it is responsible for sampling training instances from the input network structure. Prior to sampling, it should choose the source of raw features that helps preserve the expected network properties.  These properties may optionally be inferred from a specific learning task. The source of raw features can be classified into local structure (e.g., node degree, node neighbors, etc.) and global structure (multi-hop neighborhoods, node rank, etc. ) with respect to every node in the raw input graph. Different sampling methods are used for extracting features from different structures. 
The third stage is the learning node embeddings over the training set. Different embedding models can be leveraged to learn hidden features for node embedding, such as matrix factorization, probabilistic model, graph neural networks. These representative embedding models are often coupled with optimization techniques, such as hierarchical softmax, negative sampling, and attention mechanism, for better embedding effects.

\subsection{Network Data Preprocessing}
The network processing method is the data preparation stage for network representation learning. When end-users have different applications in mind for deploying NRL models, different data preprocessing methods should be employed. Hence, specific learning tasks should be discussed in the section on network data preprocessing. For example, when NRL trained models are used for node classification, if the node classification or node clustering aims to categorize new nodes based on their node state information and node edge neighbor information, then the preprocessing stage should employ techniques that can preprocess the raw graph input to obtain those required node state properties and node linkage properties for deep feature extraction (the second stage) before entering the NRL model training, the third stage of the network representation learning workflow. However, if the end-users prefer to perform node clustering or classification based on only network topology and traversal patterns over the network structure rather than node state information, then the pairwise node relationships over the entire network and their hop counts are critical in the preprocessing stage in order to learn the node distance features in terms of graph traversal semantics in the stage 2. These two steps will ensure that the NRL model training in stage 3 will deliver a high quality NRL model for end-users to perform their task specific node classification or node clustering, which are network data and domain specific in real world applications.
	

\begin{figure}[t]
	\centering
	\includegraphics[width=\linewidth]{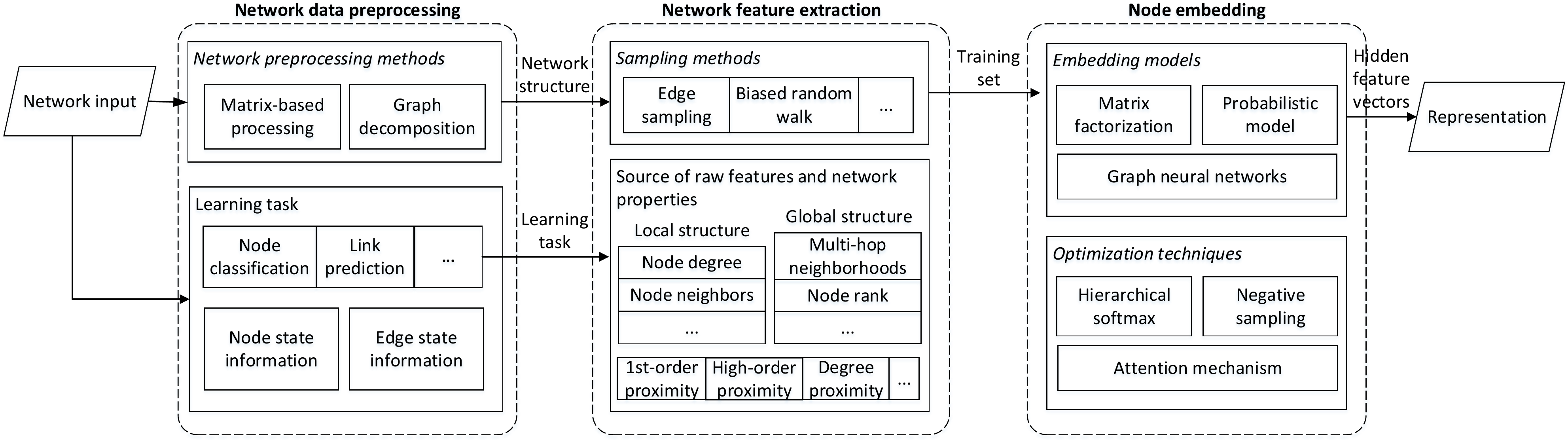}
	\caption{A unified reference framework for NRL.}
	\Description{NRL framework}
\end{figure}

\subsubsection{Network Preprocessing Methods}
To effectively capture useful features, the network structure is usually preprocessed before feature extraction. We categorize current preprocessing methods into two types: 

\textbf{Matrix-based Processing}	
In most cases, we are using an adjacency matrix \textbf{A} to represent a network $G$, where its entries could directly describe the connection between arbitrary two nodes and the connection is also the basic unit of network structure. According to the hypothesis in DeepWalk that nodes with similar contexts are similar, a node's contexts are defined as the set of nodes arrived. In order to reflect transitions between nodes, a transition matrix $\bf{P} = \bf{D^{ - 1}A}$ is proposed in \cite{DBLP:conf/cikm/CaoLX15}, where \textbf{D} is the diagonal matrix such that $\bf{D_{ii}} = \sum\nolimits_j {{A_{ij}}}$, and $\bf{P}_{ij}$ refers to the one-step transition probability from $\rm{v_i}$ to $\rm{v_j}$. Accordingly, the transition matrix can be generalized to high steps, 
and the transition matrix within $k$ steps \cite{DBLP:conf/wsdm/QiuDMLWT18} is obtained by $\bf{P^*} = \frac{1}{k}\sum\limits_{m = 1}^k {\bf{P^{m}}} $.
Yang et al. \cite{DBLP:conf/icpr/YangY18} defined two proximity matrices: the first-order proximity matrix $\bf{X}^{(1)}$ is the adjacency matrix, and the second-order proximity matrix $\bf{X}^{(2)}$ consists of $\bf{X}_{ij}^{(2)} = \cos (\bf{X}_i^{(1)},\bf{X}_j^{(1)})$, where $\bf{X}_i^{(1)}$, $\bf{X}_j^{(1)}$ are the corresponding rows of $\bf{X}^{(1)}$.
As an important branch of embedding, spectral methods require to convert adjacency matrix \textbf{A} into Laplacian matrix \textbf{L} before the Laplacian Eigenmaps \cite{DBLP:conf/nips/BelkinN01}, where $\bf{L}=\bf{D - A}$. 

\textbf{Graph Decomposition}
Graph decomposition is another important type of methods for data preprocessing, by which the original network is decomposed into multiple graphs and each of them consists of a subset of nodes and a group of edges that correspond to connections in the original network or are connected based on certain rules. These graphs may be connected together to form a new network in order to extract features for a specific network property.

To capture the structural proximity, the context graph \cite{DBLP:conf/kdd/RibeiroSF17} is proposed by leveraging the decomposition idea. The context graph is a multi-layer weighted graph $G_c$ and each layer is a complete graph of all nodes. In particular, the edges in layer $k$ are linked by nodes that are $k$-hop away from each other, and the edge weight in the layer is defined as ${w_k}(i,j) = {e^{ - {f_k}(i,j)}}$, where ${{f_k}(i,j)}$ represents the $k$-structural distance between node $\rm{v_i}$ and $\rm{v_j}$ calculated based on their $k$-hop neighborhoods. For adjacent layers, the corresponding nodes are connected by directed edges, and the edge weights in both directions are defined as ${w_{k,k + 1}}(\rm{v}) = \log ({\Gamma _k}(\rm{v}) + e)$ and ${w_{k,k - 1}}(\rm{v}) = 1$, respectively, where ${\Gamma _k}(\rm{v})$ denotes the number of edges incident to node $\rm v$ such that their weights are higher than the layer $k$'s average weight.

In social networks, ego-network $G[N_{\rm u}]$ induced on user $\rm u$'s neighborhood $N_{\rm u}$ is often used to denote her/his social circle. If the interactions of every node in the original network $G$ are divided into couples of semantic subgroups, these subgroups will capture different components of user's network behavior. Based on the idea, Epasto et al. \cite{DBLP:conf/www/EpastoP19} propose a method to convert $G$ into its persona graph $G_p$, where each original node corresponds to multiple personas. Formally, given the original network $G=\{V, E\}$ and a clustering algorithm $\mathbb{A}$, the method consists of three steps: 1) For each node $\rm{u} \in V$, its ego-network $G[N_{\rm u}]$ is partitioned into $t_{\rm u}$ disjoint components $N_{\rm u}^k$ via $\mathbb{A}$, denoted by $\mathbb{A}(G([{N_{\rm u}}])) = \{ N_{\rm u}^k|k \in [1,{t_{\rm u}}]\} $; 2) Collect a set $V_p$ of personas, where each original node ${\rm u}$ induces $t_{\rm u}$ personas; 3) Add connections between personas if and only if $(\rm u, v)$ $\in E$, $\rm u \in$ $N_{\rm v}^i$ and $\rm v$ $\in N_{\rm u}^j$. Social users often participate in different communities, the persona graph obtained via the above procedure presents a much better community structure for further embeddings.

For very large graphs whose scales exceed the capability of single machine, they must be decomposed into multiple partitions before further execution. Adam et al. \cite{DBLP:sysml/abs-1903-12287} propose a block decomposition method that first splits entity nodes into $P$ parts and then divides edges into buckets based on their source and destination entity nodes' partitions. For example, for an edge (u, v), if source u and destination v are in partitions $p_1$ and $p_2$, respectively, then it should be placed into bucket $(p_1, p_2)$. For each bucket $(p_i, p_j)$, source and destination partitions are swapped from disk, respectively, and the edges are loaded accordingly for training. To ensure that embeddings in all partitions are aligned in the same latent space, an 'inside-out' ordering is adopted to require that each bucket has at least one previously trained embedding partition.

\subsubsection{Learning task}
Node classification aims to assign each node to a suitable group such that nodes in a group have some similar features. 
Link prediction aims to find out pairs of nodes that are most likely to be connected. Node classification and link prediction are two most basic fundamental tasks for network analytics. Both tasks could be further instantiated into many practical applications such as social recommendation \cite{DBLP:conf/www/Fan0LHZTY19}, knowledge completion \cite{DBLP:conf/aaai/LinLSLZ15}, disease-gene association identification \cite{DBLP:conf/kdd/HanYZSLZ0K19}, etc. Therefore, we mainly focus on the two basic tasks. For both tasks, node's topological structure is the important basis for classification and prediction. For example, nodes with more common neighbors often have higher probability to be assigned to the same group or to be connected. This type of structure can be generalized to multi-hop neighborhoods \cite{DBLP:conf/nips/ZhangC18}, which requires to compute the transition matrix of the network.
 
In addition to pure structural information, there are other useful information available for learning tasks, e.g., node state information and edge state information.
In many real-world networks, node itself may contain some state information such as node attribute, node label, etc., and this information may be essential for some tasks. For example, in social networks, besides embedding social connections, we can also encode user attributes to obtain more comprehensive representations for individuals \cite{DBLP:conf/ijcai/ZhangYBZYZE018,DBLP:journals/tkde/LiaoHZC18}. Node attributes are still an important source of features to be aggregated for inductive embeddings \cite{DBLP:conf/nips/HamiltonYL17,DBLP:conf/iclr/BojchevskiG18}. Nodes with similar attributes and/or structures are more likely to be connected or classified together. Meanwhile, some node labels are usually fed into supervised models to boost the task of node classification \cite{DBLP:conf/iclr/KipfW17,DBLP:conf/aaai/ZhangPCU19}.
As the most common edge state information, edge weights can be integrated with topological structure to achieve more accurate classification and prediction. 
Besides, edges in some networks may have signs. Take Epinions and Slashdot as examples, users in these two social  network sites are allowed to connect to other users with either positive or negative edges, where positive edges represent trust and like while negative edges convey distrust and dislike. For link prediction on such a signed network, we have to predict not only possible edges but also signs of those edges \cite{DBLP:conf/sdm/WangTACL17,DBLP:conf/www/KimPLK18}.

\subsection{Network Feature Extraction}
The main task of NRL is to find out the hidden network features and encode such features as node embedding vectors in a low-dimensional space. 
Network properties are used to analyze and compare different network models. For a NRL task, the learned hidden features should preserve network properties so that advanced network analytics can be performed accurately and at the same time the original network properties can be preserved. For example, nodes that are closer in the original network should also be closer in the latent representation space in which the node embedding vectors are defined. Most of the NRL methods focus on preserving topological structures of nodes, such as in-degree or out-degree neighbors (degree proximity), first order proximity, random walk distance, and so forth. We categorize the node structural properties into local structure and global structure. 

Embeddings from local structure focus on the preservation of local network properties such as degree proximity, first-order proximity, etc. In comparison, global structure provides rich choices of sources of raw features to be extracted so as to preserve even more network properties.
The classification of source of raw features as well as network properties and sampling methods are summarized in Table 2.
\subsubsection{Local Structure Extraction}
Local structure reflects a node's local view about the network, which includes node degree (in-degree and out-degree), neighbors (in-degree neighbors and out-degree neighbors), node state and adjacent edge (in-edge and out-edge) state. To preserve degree proximity, Leonardo et al. \cite{DBLP:conf/kdd/RibeiroSF17} define a proximity function ${{f_k}(i,j)}$ between two nodes where each node's neighbors are sorted based on their degrees and the proximity is measured by the distance between the sorted degree sequences. The first-order proximity \cite{DBLP:conf/www/TangQWZYM15} assumes nodes that are neighbors to each other (i.e., connected via an edge) are similar in vector space, while the degree of similarity may depend on the edge state. For example, in a signed social network, neighbors with positive and negative edges are often called friends and foes, respectively. From the perspective of social psychology \cite{DBLP:conf/sdm/WangTACL17,DBLP:conf/www/KimPLK18}, when we take edge sign into account, nodes should be more similar to its friends than its foes in the representation space.
William et al. \cite{DBLP:conf/nips/HamiltonYL17} present an inductive representation learning by aggregating features extracted from neighboring node states.

\textbf{Sampling Methods} For source of raw features like degree and neighbors, training instances can be calculated or fetched directly from adjacency matrix. For adjacent edges, training instances can be generated by edge sampling. The simple edge sampling is also to fetch entries from adjacency matrix. When applied to weighted networks where the pairwise proximity has close relationship with edge weights, if edge weights have a high variance, the learning model will suffer from gradient explosion or disappearance. To address the problem, LINE \cite{DBLP:conf/www/TangQWZYM15} designs an optimized edge sampling method that fetches edges with the probabilities proportional to their weights. For node state like node attribute, its features can be extracted by leveraging existing embedding techniques, e.g., Word2vec \cite{DBLP:conf/nips/MikolovSCCD13}. If node state is given as a node label, it usually works as the supervised item to train the embedding model.

\subsubsection{Global Structure Extraction}
Structures that transcend local views can be considered global such as multi-hop neighborhoods, community, connectivity pattern, etc.
Considering that the first-order proximity matrix may not be dense enough to model the pairwise proximity between nodes, as an global view \cite{DBLP:conf/cikm/CaoLX15}, the pairwise proximity is generalized to high-order form by using $k$-step transition matrix, i.e., $\bf{P}^k$. Instead of preserving a fixed high-order proximity, Zhang et al. \cite{DBLP:conf/kdd/ZhangCWPY018} define the arbitrary-order proximity by calculating the weighted sum of all $k$-order proximities. 
Community structure is an important network property with dense intra-community connections and sparse inter-community connections, and it has been observed in many domain-specific networks, e.g., social networks, co-authoring networks, language networks, etc. Wang et al. \cite{DBLP:conf/aaai/WangCWP0Y17} introduce a community representation matrix by means of the modularity-driven community detection method, and use it to enable each node's representation similar to the representation of its community. Considering that a node may belong to multiple communities, Sun et al. \cite{DBLP:conf/cikm/SunSGOC17} define a $n \times m$ basis matrix $\bf{W}$ to reveal nodes' community memberships,  where $n$ is the node set size, $m$ is the total number of communities and $\bf{W}_{ij}$ indicates the propensity of node $\rm{v_i}$ to community $C_j$. The basis matrix is learned and preserved during the representation learning process.

Network nodes usually act as various structural roles \cite{DBLP:conf/kdd/GroverL16,DBLP:journals/corr/abs-1802-02896} and appear different connectivity patterns such as hubs, star-edge nodes, bridges connecting different clusters, etc. 
Node role proximity assumes nodes with similar roles have similar vector representations and it is a global structural property that is different from community structure, since it primarily focuses on the connection patterns between nodes and their neighbors. 


As a global structural property, node rank is always used to denote a node's importance in the network. PageRank \cite{Page98thepagerank} is a well-known approach to evaluate the rank of a node by means of its connections to others. Specifically, the ranking score of a node is measured by the probability of visiting it, while the probability is obtained from the ranking score accumulated from its direct predecessors weighted by the reciprocal of its out-degree. Lai et al. \cite{DBLP:conf/nips/LaiHCYL17} demonstrate that node representations with global ranking preserved can potentially improve both results of ranking-oriented tasks and classification tasks.
Node degree distribution is also a global inherent network property. For example, scale-free property refers to a fact that node degrees follow a heavy-tailed distribution, and it has proven to be ubiquitous across many real networks, e.g., Internet, social networks, etc. The representation learning for scale-free is explored in \cite{DBLP:conf/aaai/Feng0H0Z18}.

Another global structural property is proposed in Struc2vec \cite{DBLP:conf/kdd/RibeiroSF17} that considers structural similarity from network equivalence perspective without requiring two nodes being nearby, i.e., independent of nodes' network positions. To reflect this property, Struc2vec presents a notion of node structural identity that refers to a node's global sense. Struc2vec uses the multi-layer graph output from data preprocessing stage to measure node similarity at different scales. In the bottom layer, the similarity exclusively  depends on node degrees. In the top layer, the similarity lies in the entire network.
Tu et al. \cite{DBLP:conf/kdd/TuCWY018} propose a similar concept, i.e., regular equivalence, to describe the similarity between nodes that may not be directly connected or not having common neighbors. According to its recursive definition, neighbors of regularly equivalent nodes are also regularly equivalent. To ensure the property of regular equivalence, each node's embedding is approximated by aggregating its neighbors' embeddings. After updating the learned representations iteratively, the final node embedding is capable of preserving the property in a global sense.

Different types of global structure are used to reflect different network properties where multi-hop neighborhoods, node connectivity pattern and node identity are used to preserve pairwise proximity, which reflects a pairwise relationship between nodes, including high-order proximity, node role proximity, and node identity proximity. For example, the node community membership reflects a relationship between a node and a group of nodes, which share some common network properties. Furthermore, node rank and node degree distribution are used to preserve a kind of distribution-based network property, including node importance ranking, or a relationship between a node and the entire network, such as the scale free network whose degree distribution follows a power law.

\textbf{Sampling Methods} For source of raw features like multi-hop neighborhoods, they can be obtained by matrix power operation, i.e., $\bf{A}^k$, but the computation suffers from high complexity.
Random walk and its variants are widely explored to capture the desirable network properties with high confidence. For example, DeepWalk \cite{DBLP:conf/kdd/PerozziAS14} presents a truncated random walk to generate node traveling paths. It uses co-occurrence frequencies between node and its multi-hop neighborhoods along these paths to reflect their similarity and capture the high-order proximity accordingly. 

From the perspective of community structure, due the dense intra-community connections, nodes within the same community have higher probability to co-occur on the traveling paths than nodes in different communities. Hence random walk can also be used to capture the community structure. When we consider the hierarchy of communities, different communities may have different scales. The regular random walk makes the training set having more entries from $\bf{A}^i$ than from $\bf{A}^j$ ($1 \le i < j$), and then it is biased towards preserving small-scale community structure. Walklets \cite{DBLP:conf/asunam/PerozziKCS17} presents a skipped random walk to sample multi-scale node pairs by skipping over steps in each traveling path.

Another drawback of random walk is that it requires too many steps or restarts to cover a node's neighborhoods. To improve its coverage, Diff2Vec \cite{DBLP:journals/corr/abs-2001-07463} present a diffusion-based node sequence generating method that consists of two steps: 1) Diffusion graph generation, which is in charge of generating a diffusion graph $DG_i$ for each node $\rm{v_i}$. $DG_i$ is initialized with {$\rm{v_i}$}, and then randomly fetch node $\rm{v_j}$ from $DG_i$ and node $\rm{v_k}$ from $\rm{v_j}$'s neighborhoods in the original graph, append two nodes and the edge $e_{jk}$ to $DG_i$. The above process is repeated until $DG_i$ grows to the predefined size. 2) Node sequence sampling, which generates Euler walk from $DG_i$ as the node sequence. To make sure $DG_i$ is Eulerian, $DG_i$ is converted to a multi-graph by doubling every edge into two edges.

Real-world networks often exhibit a mixture of multiple network properties. In order to capture both community structure and node role proximity, node2vec \cite{DBLP:conf/kdd/GroverL16} designs a flexible biased random walk that generates traveling paths in an integrated fashion of BF (breadth-first) sampling and DF (depth-first) sampling. To this end, two parameters $p$ and $q$ are introduced to smoothly interpolate between two sampling methods, where $p$ decides the probability of re-fetching a node in the path while $q$ allows the sampling to discriminate between inward and outward nodes.

In scale-free networks, a tiny fraction of "big hubs" usually attracts most edges. Considering that connecting to "big hubs" does not imply proximity as strong as connecting to nodes with mediocre degrees, a degree penalty based random walk \cite{DBLP:conf/aaai/Feng0H0Z18} is proposed. For a pair of connected nodes ($\rm v_i$, $\rm v_j$), its principle is to reduce the likelihood of $\rm v_j$ being sampled as $\rm v_i$'s context when $\rm v_i$ has a high degree and they do not share many common neighbors. To this end, the jumping probability from $\rm v_i$ to $\rm v_j$ is defined as $\frac{{{C_{ij}}}}{{{{({D_{ii}}{D_{jj}})}^\beta }}}$, where $C_{ij}$ denotes the first and second order of proximity between two nodes, $D_{ii}$ and $D_{jj}$ are their degrees, and $\beta$ is a parameter.

As an anonymous version of random walk, anonymous walk \cite{DBLP:conf/icml/IvanovB18} provides a flexible way to reconstruct a network. For a random walk $\rm rw=(v_1, v_2, ..., v_k)$, its corresponding anonymous walk is the sequence of node's first occurrence positions, i.e., $\rm aw=(f(v_1), f(v_2), ..., f(v_k))$, where $\rm f({v_i}) = \mathop {\min }\limits_{{p_j} \in pos(rw,v_i)} pos(rw,v_i)$. For an arbitrary node $\rm v \in G$, a known distribution ${\rm D}_l$ over anonymous walks of length $l$ is sufficient to reconstruct a subgraph of $\rm G$ that is induced by all nodes located within $r$ hops away from $\rm v$. Therefore, anonymous walk can be used to preserve global network properties like high-order proximity and community structure by approximating the distributions.

One of the baseline approaches to extracting global structure is to use random walk as the sampling method. For complex types of global structure, e.g., multi-hop neighborhoods and node community membership, an integrated sampling method is often recommended, which combines random walk with other types of graph traversal methods, such as anonymous walk. An advantage of using anonymous walk as the sampling approach is that it is sufficient to reconstruct the topology around a node by utilizing distribution of anonymous walks of a single node, because anonymous walk captures richer semantics than random walk.

\begin{table}[t]
	\centering
	\tiny
	\setlength{\abovecaptionskip}{2pt}
	\caption{Classification of source of raw features, network property and sampling method}
	\begin{tabular}{|c|l|l|l|}
		\hline
		\textbf{Categority} & \textbf{Source of raw features} & \textbf{Network property} & \textbf{Sampling method} \\
		\hline
		\multicolumn{1}{|c|}{\multirow{4}[1]{*}{Local structure}} & Degree/in-degree/out-degree \cite{DBLP:conf/kdd/RibeiroSF17} & Degree proximity & Directly from adjacency matrix \\
		\cline{2-4}    \multicolumn{1}{|c|}{} & Neighbors/in-degree neighbors/out-degree neighbors \cite{DBLP:conf/www/TangQWZYM15} & First-order proximity & Directly from adjacency matrix \\
		\cline{2-4}    \multicolumn{1}{|c|}{} & Node state \cite{DBLP:conf/nips/HamiltonYL17}& Node state proximity & Directly from node state information \\
		\cline{2-4}    \multicolumn{1}{|c|}{} & Adjacent edge state \cite{DBLP:conf/sdm/WangTACL17,DBLP:conf/www/KimPLK18}& First-order proximity & Weighted edge sampling \\
		\hline
		\multicolumn{1}{|c|}{\multirow{6}[0]{*}{Global structure}} & Multi-hop neighborhoods \cite{DBLP:conf/kdd/ZhangCWPY018}& High-order proximity & Random/Anonymous walk \\
		\cline{2-4}    \multicolumn{1}{|c|}{} & Node community membership \cite{DBLP:conf/aaai/WangCWP0Y17,DBLP:conf/cikm/SunSGOC17}& Community structure & Random/Anonymous walk \\
		\cline{2-4}    \multicolumn{1}{|c|}{} & Node connectivity pattern  \cite{DBLP:conf/kdd/GroverL16,DBLP:journals/corr/abs-1802-02896}& Node role proximity & Random walk \\
		\cline{2-4}    \multicolumn{1}{|c|}{} & Node degree distribution \cite{DBLP:conf/aaai/Feng0H0Z18}& Scale-free & Degree penalty based random walk \\
		\cline{2-4}    \multicolumn{1}{|c|}{} & Node rank \cite{DBLP:conf/nips/LaiHCYL17} & Node importance ranking & Random walk \\
		\cline{2-4}    \multicolumn{1}{|c|}{} & Node identity \cite{DBLP:conf/kdd/RibeiroSF17,DBLP:conf/kdd/TuCWY018} & Node identity proximity & Random walk \\
		\hline
	\end{tabular}%
	\label{tab2}%
\end{table}%



\subsection{Node Embedding}
Recent years many efforts have been devoted to the design of node embedding model. We are trying to review those work from a universal perspective of autoencoder. In general, node embedding is equivalent to an optimization problem that encodes network nodes into latent vectors by means of an encoding function $\Phi :V \to {\mathbb{R}^d}$. 
Meanwhile the objective is to ensure that the results decoded from vectors preserve the network properties we intend to incorporate, where the decoder is represented by a function of encoding results, i.e., $\Psi :{[{\mathbb{R}^d}]^t} \to \mathbb{R}$, where $t$ denotes the number of input arguments.
The output vectors are the latent representations of hidden features learned by $\Phi$, and they are also the expected node representations. 
\subsubsection{Embedding Models} We classify the current embedding models into the following three types.

\textbf{Matrix Factorization}
Given the matrix of input network, matrix factorization embedding factorizes the matrix to get a low-dimensional matrix as the output collection of node representations. Its basic idea can be traced back to the matrix dimensionality reduction techniques \cite{DBLP:journals/corr/abs-1808-02590}. According to the matrix type, we categorize the current work into relational matrix factorization and spectral model.

\emph{(1) Relational Matrix Factorization} Matrix analysis often requires figuring out the inherent structure of a matrix by a fraction of its entries, and it is always based on an assumption that a matrix $\bf{M}_{n \times n}$ admits an approximation of low rank $d \ll n$. Under this assumption, the objective of matrix factorization corresponds to finding a matrix $\bf{W}_{d \times n}$ such that $\bf{W}^T \bf{W}$ approximates $\bf{M}$ with the lowest loss $L$, where $L:{[\mathbb{R}]^2} \to \mathbb{R}$ is a user-defined loss function. In the autoencoder paradigm, the encoder is defined by the matrix $\bf{W}_{d \times n}$, e.g., $\Phi (\rm{v_i}) = \bf{w}_i$, each column $\bf{w}_i$ represents a vector. The decoder is defined by the inner product of two node vectors, e.g., $\Psi (\bf{w}_i, \bf{w}_j) = \bf{w}_i^T \bf{w}_j$, so as to infer the reconstruction of the proximity between node $\rm{v_i}$ and $\rm{v_j}$. When we focus on preserving the first-order proximity $\mu$, i.e., $\mu(\rm{v_i}, \rm{v_j})=\bf{A}_{ij}$, and let $\Omega$ be the training set sampled from $\bf{A}$, then the objective is to find $\bf{W}$ to minimize the reconstruction loss, i.e.,
\begin{equation}\label{9}
\mathop {\arg \min }\limits_{\bf{W}} L = \sum\limits_{({\rm{v_i}},{\rm{v_j}}) \in \Omega } {\left\| {\Psi (\bf{w}_i,\bf{w}_j) - \mu ({v_i},{v_j})} \right\|_F^2}  = \left\| {\bf{W}^T\bf{W} - A} \right\|_F^2 ,
\end{equation}
where Frobenius norm $\left\|  \cdot  \right\|_F^2$ is often used to define the loss function. Singular value decomposition (SVD) \cite{DBLP:journals/nm/golub} is a well-known matrix factorization approach that can find the optimal rank $d$ approximation of proximity matrix $\bf{A}$. If the network property focuses on high-order proximity, the proximity matrix can be replaced by power matrix, e.g., $\bf{A}^k$, $\bf{P}^k$, $k>1$. For example, GraRep  \cite{DBLP:conf/cikm/CaoLX15} implements node embedding by factorizing $\bf{P}^k$ into two matrices $\bf{W}$ and $\bf{M}$, 
\begin{equation}\label{10}
\mathop {\arg \min }\limits_{\bf{W}} \left\| {\bf{W^T}\bf{M} - {P^k}} \right\|_F^2,
\end{equation}
where $\bf{W}$ denotes the representation matrix and $\bf{M}$ denotes the parameter matrix in decoder, e.g., the unary decoder $\Psi (\bf{w_i}) = \bf{w}_i^TM$ computes the reconstructed proximities between $\rm{v_i}$ and other nodes.

In the autoencoder paradigm, the equivalence between DeepWalk and matrix factorization can be proved by making the following analogies: 1) Define the pairwise proximity $\mu(\rm{v_i}, \rm{v_j})$ as the co-occurrence probability inferred from the training set.
2) Let $\bf{W}$ and $\bf{M}$ corresponds to $\bf{W}^T$ and $\bf{M^T}$ in DeepWalk, where each column $\bf{w}_i$, $\bf{m}_i$ of them refers to the representation of node $\rm{v_i}$ acting as target node and context node, respectively.

In addition, matrix factorization can also incorporate additional information into node embeddings. For example, given a text feature matrix $\bf{T}_{x \times n}$ where $n$ denotes the node set size and $x$ denotes the feature vector dimension, 
TADW \cite{DBLP:conf/ijcai/YangLZSC15} applies inductive matrix completion to the node embedding and defines the following matrix factorization problem:
\begin{equation}\label{11}
\mathop {\arg \min }\limits_{{\bf{W}},{\bf{M}}} \left\| {{{\bf{P}}^{\bf{*}}} - {{\bf{W}}^{\rm{T}}}{\bf{MT}}} \right\|_F^2 + \frac{\lambda }{2}\left( {\left\| {\bf{W}} \right\|_F^2 + \left\| {\bf{M}} \right\|_F^2} \right),
\end{equation} 
where $\lambda$ is a harmonic factor. The output matrices $\bf{W}$ and $\bf{MT}$ factorized from the transition matrix $\bf{P}^*$ can be regarded as the collection of node embeddings and concatenated as a $n \times 2d$ representation matrix.

\emph{(2) Spectral Model}
In spectral model, a network is mathematically represented by a Laplacian matrix, i.e., $\bf{L} =D -A$, where adjacency matrix $\bf{A}$ acts as the proximity matrix, and the entry $\bf{D}_{ii}$ of diagonal matrix $\bf{D}$ describes the importance of node $\rm{v}_i$.  $\bf{L}$ is a symmetric positive semidefinite matrix that can be thought of as an operator on functions defined on the original network. Let $\bf{W}_{n \times d}$ be the representation matrix, it also acts as the encoder that maps the network to a $d$-dimensional space. The representations should assure neighboring nodes stay as close as possible. As a result, the decoder can be defined as $\bf{w}_i^TLw_i$ according to Laplacian Eigenmaps \cite{DBLP:journals/neco/BelkinN03}, where $\bf{w}_i$ is the $i$th column of $\bf{W}$. Then the node embedding problem is defined as follows:
\begin{equation}\label{12}
\mathop {\arg \min }\limits_{{{\bf{W}}^{\rm{T}}}{\bf{DW}} = {\bf{I}}} {{\bf{W}}^{\rm{T}}}{\bf{LW}},
\end{equation}
where ${{{\bf{W}}^{\rm{T}}}{\bf{DW}} = {\bf{I}}}$ is imposed as a constraint to prevent collapse onto a subspace of dimension less than $d-1$. The solution $\bf{W}$ consists of the eigenvectors $\bf{w_i}$ corresponding to the lowest $d$ eigenvalues of the generalized eigenvalue problem $\bf{L}w = \lambda D w$. Furthermore, given an additional information matrix $\bf{X}$ describing node attribute features, the node embedding can be derived by following the idea of locality preserving projection (LPP) \cite{DBLP:conf/nips/HeN03} that introduces a transformation matrix $\bf{M}$ to realize the mapping $\bf{w}_i=M^T x_i$, where $\bf{w}_i$ and $\bf{x}_i$ are the $i$th column of $\bf{W}$ and $\bf{X}$. Similarly, the solution $\bf{M}$ is obtained by computing the mapping vectors $\bf{m}$ corresponding to $d$ lowest eigenvalues of the problem $\bf{XLX^T}m = \lambda XDX^T m$, and the node embedding incorporates additional information by $\bf{W}=M^T X$.

\textbf{Probabilistic Model}
Probabilistic model is specifically designed for node embedding via preserving the pairwise proximity measured by a flexible probabilistic manner. 

\emph{(1) Skip-gram Model}
{Skip-gram} is the classic embedding model that converts pairwise proximity to the conditional probability between node and its context. For example, DeepWalk \cite{DBLP:conf/kdd/PerozziAS14} relies on the co-occurrence probability derived from random traveling paths to preserve the second-order proximity during node embedding. 

\emph{(2) Edge Probabilistic Model}
Edge probabilistic model enforces node embeddings designed primarily for network reconstruction. For example, LINE \cite{DBLP:conf/www/TangQWZYM15} relies on the idea of edge probability reconstruction without assistance of random walk sampling. It focuses on preserving both first-order and second-order proximities by defining two decoders, i.e.,
\begin{equation}\label{13}
{\Psi _1}({{\bf{v}}_{\bf{i}}},{{\bf{v}}_{\bf{j}}}) = \Pr (\rm{v_i},\rm{v_j}) = \frac{1}{{1 + \exp ( - {{\bf{v}}_{\bf{i}}}^{\rm{T}}{{\bf{v}}_{\bf{j}}})}}, \, {\Psi _2}({{\bf{v}}_{\bf{i}}},{{\bf{m}}_{\bf{j}}}) = \Pr ({v_j}|{v_i}) = \frac{{\exp ({{\bf{m}}_{\bf{j}}}^{\rm{T}}{{\bf{v}}_{\bf{i}}})}}{{\sum\nolimits_{k = 1}^n {\exp ({{\bf{m}}_{\bf{k}}}^{\rm{T}}{{\bf{v}}_{\bf{i}}})} }},
\end{equation}
where $\bf m_k$ refers to the representation of $\rm v_k$ acting as a context node. The objectives to be optimized are based on the loss functions derived from the distance $Dis(\cdot,\cdot)$ between two distributions, i.e.,
\begin{equation}\label{14}
\mathop {\arg \min }\limits_{{{\bf{v}}_{\bf{i}}}} \sum\limits_{i = 1}^n {{\lambda _i}Dis({\Psi _1}({{\bf{v}}_{\bf{i}}}, \cdot ),{{\hat \Psi }_1}({{\bf{v}}_{\bf{i}}}, \cdot ))} , \, \mathop {\arg \min }\limits_{{{\bf{v}}_{\bf{i}}}} \sum\limits_{i = 1}^n {{\lambda _i}Dis({\Psi _2}({{\bf{v}}_{\bf{i}}}, \cdot ),{{\hat \Psi }_2}({{\bf{v}}_{\bf{i}}}, \cdot ))} 
\end{equation}
where ${{\hat \Psi }_1}({{\bf{v}}_{\bf{i}}},{{\bf{v}}_{\bf{j}}}) = \frac{{{s_{ij}}}}{{\sum\nolimits_{{e_{xy}} \in E} {{w_{xy}}} }},{{\hat \Psi }_2}({{\bf{v}}_{\bf{i}}},{{\bf{v}}_{\bf{j}}}) = \frac{{{s_{ij}}}}{{\sum\nolimits_{{e_{ik}} \in E} {{s_{ik}}} }}$ are the empirical probabilities and $s_{ij}$ is the weight of edge $e_{ij}$.


\textbf{Graph Neural Networks}
It is not difficult to find that each node's embedding via the above models requires the participation of its neighboring nodes, e.g., building $k$-order proximity matrix with nodes and their $k$-step neighborhoods in matrix factorization model, and counting node's co-occurrence frequency with its context nodes in skip-gram model. The common idea can be intuitively generalized to a more general model, graph neural networks (GNNs) that follow a recursive neighborhood aggregation or message passing scheme. 	
Graph convolutional network (GCN) \cite{DBLP:journals/corr/BrunaZSL13} is a very popular variant of GNNs, where each node's representation is generated by a convolution operation that aggregates its own features and its neighboring nodes' features. Graph isomorphism network (GIN) \cite{DBLP:conf/iclr/XuHLJ19} is another recently proposed variant with a relatively strong expressive power.

\emph{(1) Graph Convolutional Network}
In terms of the definition of convolution operation \cite{DBLP:journals/corr/abs-1901-00596}, GCNs are often grouped into two categories: Spectral GCNs and Spatial GCNs. Spectral GCNs \cite{DBLP:conf/nips/DefferrardBV16,DBLP:journals/corr/BrunaZSL13} define the convolution as conducting the eigendecomposition of the normalized Laplacian matrix ${\bf{L}} = {{\bf I}_{\rm{n}}} - {{\bf{D}}^{{\rm{ - }}\frac{{\rm{1}}}{{\rm{2}}}}}{\bf{A}}{{\bf{D}}^{{\rm{ - }}\frac{{\rm{1}}}{{\rm{2}}}}}$ in Fourier domain. An intuitive way to explain spectral convolution is to regard it as an operation that uses a filter $\bf{g}$ to remove noise from a graph signal $\bf{x} \in \mathbb{R}^n$. The graph signal denotes a feature vector of the graph with entry $\bf{x}_i$ representing node $\rm{v}_i$'s value.
Let $\bf{U}$ be the matrix of eigenvectors of $\bf{L}$,
graph Fourier transform is defined as a function that projects input signal $\bf{x}$ to the orthonormal space formed by $\bf{U}$, i.e.,  $\mathcal{F}(\bf{x})=\bf{U}^Tx$. Let $\bf{\hat{x}}$ be the transformed signal, entries of $\bf{\hat{x}}$ correspond to the coordinates of the input signal in the new generated space, and the inverse Fourier transform is defined as $\mathcal{F}^{-1}(\bf{\hat{x}})=\bf{U}\hat{x}$. When the filter $\bf{g}$ is parameterized by $\bf \theta \in \mathbb{R}^n$ and defined as ${{\bf{g}}_{\bf{\theta }}} = diag({{\bf{U}}^{\rm{T}}}{\bf{g}})$, then the convolution operation against signal $\bf{x}$ with filter $\bf{g}_\theta$ is represented by
\begin{equation}\label{15}
{{\bf{g}}_{\bf{\theta }}} \star {\bf{x}} = {{\mathcal{F}}^{{\bf{ - 1}}}}(\mathcal{F}({\bf{g}}) \circ \mathcal{F}({\bf{x}})) = {\bf{U}}({{\bf{U}}^{\rm{T}}}{\bf{g}} \circ {{\bf{U}}^{\rm{T}}}{\bf{x}}) = {\bf{U}}{{\bf{g}}_{\bf{\theta }}}{{\bf{U}}^{\rm{T}}}{\bf{x}},
\end{equation}
where $\circ$ represents the Hadamard product. Nevertheless, The eigendecomposition of Laplacian matrix $\bf{L}$ is expensive for large graphs, and the complexity of multiplication with $\bf{U}$ is $O(n^2)$. To address the complexity problem, Hammond et al. \cite{DBLP:journals/corr/abs-0912-3848} suggests that the filter $\bf{g}_\theta$ can be approximated by an abridged expansion based on Chebyshev polynomials $T_k(\bf{x})$ up to $K$th order. The convolution is redefined as
\begin{equation}\label{16}
{{\bf{g}}_{{\bf{\theta }}'}} \star {\bf{x}} \approx \sum\limits_{k = 0}^K {{\bf{\theta }}{'_k}{T_k}({\bf{\tilde L}}){\bf{x}}} ,
\end{equation} 
where ${\bf{\tilde L}} = \frac{2}{{{\lambda _{\max }}}}{\bf{L}} - {{\bf{I}}_{\rm{n}}}$ and  $T_k(\bf{x}) = 2\bf{x}T_{k-1}(\bf{x})-T_{k-2}(\bf{x})$ with $T_{1}(\bf{x})=x$ and $T_{0}(\bf{x})=\rm 1$. The above convolution operation is $K$-localized since it requires the participation of the neighboring nodes within $K$-hop away from the central node. Spectral GCNs model consists of multiple convolution layers of the form Eq. 16 where each layer is followed by a point-wise non-linearity. From the perspective of autoencoder, each node's feature embedding is encoded by the convolution operation, so that the graph convolution actually acts as the encoder and we call it spectral convolution encoder or filter encoder here. 
In order to deal with multi-dimensional input graph signals, the spectral convolution encoder is generalized to account for the signal matrix $\bf{X} \in \mathbb{R}^{n \times c}$ with $c$ input channels and $d$ filters, i.e.,
\begin{equation}\label{17}
\Phi ({\bf{X}},{\bf{A}}) = {{{\bf{\tilde D}}}^{ - \frac{1}{2}}}{\bf{\tilde A}}{{{\bf{\tilde D}}}^{ - \frac{1}{2}}}{\bf{X\Theta }},
\end{equation}
where $\bf{\tilde{A}}=A+I_n$, $\bf{\tilde{D}}_{ii}={\sum\nolimits_j {{\bf{\tilde A}}} _{ij}}$ and ${\bf{\Theta }} \in {\mathbb{R}^{c \times d}}$ is matrix of filter parameters. The signal matrix is often initialized by the input graph information such as node attributes. It is noted that the filter parameters ${\bf{\Theta }}$ can be shared over the whole graph, which significantly decreases the amount of parameters as well as improves the efficiency of filter encoder. 

Similar to convolution neural networks on images, spatial GCNs \cite{DBLP:conf/nips/HamiltonYL17} consider graph nodes as image pixels and directly define convolution operation in the graph domain as the feature aggregation from neighboring nodes. To be specific, the convolution acts as the encoder to take the aggregation of the central node representation and its neighbors' representations to generate an updated representation, and here we call it spatial convolution encoder or aggregation encoder. In order to explore the depth and breadth of a node's receptive field, spatial GCNs usually stack multiple convolution layers. For a $K$-layer spatial GCN, its aggregation encoder ${\Phi}$ is defined as follows:
\begin{equation}\label{18}
\left\{ \begin{array}{l}
\Phi ({\bf{X}},{\bf{A}}) = NORM({{\bf{H}}^{\rm{K}}}),{{\bf{H}}^{\rm{K}}} = ({\bf{h}}_{\rm{1}}^{\rm{K}},...,{\bf{h}}_{\rm{n}}^{\rm{K}}), \\ 
{\bf{h}}_{\rm{v}}^{\rm{k}} = \sigma ({{\bf{W}}^{\rm{k}}} \cdot AG{G_{\rm{k}}}({\bf{h}}_{\rm{v}}^{{\rm{k - 1}}},\{ {\bf{h}}_{\rm{u}}^{{\rm{k - 1}}}\} )),\forall {\rm{u}} \in {N_{\rm{v}}}, \forall {\rm{k} \in [0,\rm{K}]}, \\ 
\end{array} \right.
\end{equation}
where $\bf{h}_{\rm{v}}^{\rm{k}}$ is node $\rm v$'s representation (also called hidden state in some other literatures) in the $k$th layer with $\bf{h}_{\rm{v}}^{\rm{0}}=\bf{x}_v$. 
$AGG_{\rm{k}}$ is an aggregation function responsible for assembling a node's neighborhood information in the $k$th layer, and parameter matrix $\bf{W}^k$ specifies how to do aggregation from neighborhoods, like filters in spectral GCNs, $\bf{W}^k$ is shared across all graph nodes for generating their representations.
As the layer deepens, the node representations will contain more information aggregated from wider coverage of nodes. The final node representations are output as the normalized matrix of node representations on layer $\rm K$, i.e., $NORM({\bf{H}}^{\rm{K}})$.


\begin{table}[tbp]
	\centering
	\tiny
	\setlength{\abovecaptionskip}{2pt}
	\caption{Classification of node embedding models}
	\begin{tabular}{|c|p{2cm}<{\centering}|p{2.2cm}<{\centering}|p{2cm}<{\centering}|p{1.8cm}<{\centering}|l|}
		\hline
		\textbf{Category} & \textbf{Sub-category} & \textbf{Encoder} & \textbf{Decoder} & \textbf{Optimization objective} & \textbf{Publication} \\
		\hline
		\multicolumn{1}{|c|}{\multirow{2}[10]{*}{\tabincell{c}{Matrix \\ Factorization}}} & Relational matrix factorization &  $\Phi (\rm{v_i}) = \bf{w}_i$   &  $\Psi (\bf{w}_i, \bf{w}_j) = \bf{w}_i^T \bf{w}_j$, $\Psi (\bf{w_i}) = \bf{w}_i^TM$, $\Psi (\bf{w_i}) = \bf{w}_i^TMT$  &  Eq. 9,  Eq. 10,  Eq. 11  & \cite{DBLP:conf/cikm/CaoLX15,DBLP:conf/ijcai/YangLZSC15} \\
		\cline{2-6}    \multicolumn{1}{|c|}{} & Spectral model &  $\Phi (V) = \bf{W}$, $\Phi (V) = \bf{M}^TX$  & $\bf{w}_i^TLw_i$, $\bf{x_iLx_i^Tm_i}$  & Eq. 12  & \cite{DBLP:journals/neco/BelkinN03,DBLP:conf/nips/HeN03} \\
		\hline
		\multicolumn{1}{|c|}{\multirow{2}[4]{*}{\tabincell{c}{Probabilistic \\ Model}}} & Skip-gram model &  $\Phi (\rm{v_i}) = \bf{w}_i$  &  Eq. 2 & Eq. 4  & \cite{DBLP:conf/kdd/PerozziAS14} \\
		\cline{2-6}    \multicolumn{1}{|c|}{} & Edge probabilistic model & $\Phi (\rm{v_i}) = \bf{w}_i$  &   Eq. 13  &  Eq. 14   & \cite{DBLP:conf/www/TangQWZYM15} \\
		\hline
		\multicolumn{1}{|c|}{\multirow{3}[14]{*}{\tabincell{c}{Graph Neural \\ Networks}}} & Spectral Graph Convolutional Network & Eq. 17  & $\Psi_{\bf{\theta}}(\Phi(\bf{x}_i,A))$ & Eq. 20  & \cite{DBLP:journals/corr/abs-0912-3848,DBLP:conf/iclr/KipfW17,DBLP:nips/bdl/KipfW16a}    \\
		\cline{2-6}    \multicolumn{1}{|c|}{} & Spatial Graph Convolutional Network &  Eq. 18 &  $\Psi_{\bf{\theta}}(\Phi(\bf{x}_i,A))$  &  Eq. 20  & \cite{DBLP:conf/nips/HamiltonYL17}  \\
		\cline{2-6}    \multicolumn{1}{|c|}{} & Graph Isomorphism Network & Eqs. 18, 19 &  $\Psi_{\bf{\theta}}(\Phi(\bf{x}_i,A))$  &  Eq. 20  & \cite{DBLP:conf/iclr/XuHLJ19} \\
		\hline
	\end{tabular}%
	\label{tab3}%
\end{table}%
\emph{(2) Graph Isomorphism Network}
The representational power of a GNN primarily depends on whether it maps two multisets to different representations, where multiset is a generalized concept of a set that allows multiple instances for its elements. The aggregation operation can be regarded as a class of functions over multisets that their neural networks can represent. To maximize the representational power of a GNN, its multiset functions must be injective. GCN has been proven to be not able to distinguish certain simple structural features as its aggregation operation is inherently not injective. To model injective multiset functions for the neighbor aggregation, GIN \cite{DBLP:conf/iclr/XuHLJ19} presents a theory of deep multisets that parameterizes universal multiset functions with neural networks. With the help of multi-layer perceptrons (MLPs), GIN updates node representations as follows:
	\begin{equation}\label{19}
	{\bf{h}}_{\rm{v}}^{\rm{k}} = MLP^{\rm{k}}\left( {\left( {1 + {{\varepsilon} ^{\rm{k}}}} \right){\bf{h}}_{\rm{v}}^{\rm{k-1}} + \sum\limits_{{\rm{u}} \in {N_{\rm{v}}}} {{\bf{h}}_{\rm{u}}^{\rm{k-1}}} } \right),
	\end{equation}
	where $\varepsilon$ is a learnable parameter or a fixed scalar. The aggregation encoder $\Phi$ can be defined in the same way of Eq. 18.

For both GCN and GIN, the decoder can be designed in any form of the previously discussed decoders. Recently, many efforts have been devoted to designing task-driven GCNs, so that the decoder has to incorporate supervision from a specific task. For instance, assume that each node has a label chosen from a label set $\bf{y}$. A function like sigmoid, $y_i=\Psi_{\bf{\theta}}(\Phi(\bf{x}_i,A))$, can be defined as a decoder to realize the mapping from a node representation $\Phi(\bf{x}_i,A)$ to its corresponding label $y_i$, where parameters $\theta$ are learnable throughout the embedding process. We use function $f(\hat{y_i}, \Psi_{\bf{\theta}}(\Phi(\bf{x}_i,A)))$ to measure the loss between the true label $\hat{y_i}$ and the predicted one $y_i$, then the objective is concluded as
\begin{equation}\label{key20}
\mathop {\arg \min }\limits_{\Phi ({{\bf{x}}_{\rm{i}}},{\bf{A}})} \sum\limits_{{\rm{i}} = 1}^n {f({{\hat y}_{\rm{i}}},\Psi_{\theta} (\Phi ({{\bf{x}}_{\rm{i}}},{\bf{A}})))}.
\end{equation}

We summarize the specific classification of node embedding models outlined in Table 3 and make comparisons among them as shown in Table 4. 
\begin{table}[t]
	\centering
	\tiny
	\setlength{\abovecaptionskip}{2pt}
	\caption{Comparison of node embedding models}
	\begin{tabular}{|c|p{5cm}|p{5cm}|}
		\hline
		\textbf{Category} & \multicolumn{1}{c|}{\textbf{Advantage}} & \multicolumn{1}{c|}{\textbf{Disadvantage}} \\
		\hline
		Matrix Factorization & 1) has the solid theoretical foundation available to enhance its interpretability. 
		2) additional information, e.g., node state and edge state, can be easily incorporated into the matrix factorization model. & 1) high computation cost. 
		2) hard to be applied to dynamic embedding. \\
		\hline
		Probabilistic Model & relatively efficient compared to other models especially when it was designed for network reconstruction. & hard to be applied to dynamic embedding. \\
		\hline
		Graph Neural Networks & 1) naturally supports embedding of both structural and additional information. 
		2) easy to be applied to dynamic embedding. & high computation cost. \\
		\hline
	\end{tabular}%
	\label{tab4}%
\end{table}%

\subsubsection{Optimization Techniques}
Many of the above models have achieved non-trivial embedding results. The success of these models inevitably relies on some optimization techniques that have been proposed to assist the node embedding from various aspects such as the computational complexity reduction, the acceleration of embedding process
and the enhancement of training efficiency and effectiveness. In this section, we will show several popular optimization techniques with a focus on how they work at different stages of NRL, and summarize these techniques in Table 5.

\textbf{Hierarchical Softmax} 
As a variant of softmax, hierarchical softmax \cite{DBLP:conf/aistats/MorinB05,DBLP:conf/kdd/PerozziAS14} has been proposed to speed up the training process. To this end, hierarchical softmax leverages a well-designed tree structure with multiple binary classifiers to compute the conditional probabilities for each training instance (see Sec. 2 for details). Therefore, it helps decoder reduce the complexity from $O(n)$ to $O(\log n)$.

\textbf{Negative Sampling} 
As an alternative to hierarchical softmax, noise contrastive estimation (NCE) \cite{DBLP:journals/jmlr/GutmannH12} suggests that logistic regression can be used to help models distinguish data from noise. Negative sampling (NEG) is a simplified version of NCE that was firstly leveraged by Word2vec \cite{DBLP:conf/nips/MikolovSCCD13} to help the skip-gram model to deal with the difficulty of high computational complexity w.r.t. model training. Specifically, NEG needs a noise distribution $P_n(\rm v)$ to generate $k$ negative samples for every positive one. $P_n(\rm v)$ can be arbitrarily chosen, and more often it is empirically set by raising the frequency to the $3/4$ power for the best quality of embeddings. The training loss of NEG is formally defined as
\begin{equation}\label{20}
L =  - \log \sigma ({{\bf{v}}_{\rm{O}}}^{\rm{T}}\Phi ({{\rm{v}}_{\rm{I}}})) - \sum\limits_{{{\rm{v}}_{\rm{j}}} \in {\rm{V}}_{neg}} {{E_{{{\rm{v}}_{\rm{j}}} \sim {P_{n({\rm{v}})}}}}\left[ {\log \sigma ( - {{\bf{v}}_{\rm{j}}}^{\rm{T}}\Phi ({{\rm{v}}_{\rm{I}}}))} \right]}  ,
\end{equation}
where ${\rm{V}}_{neg}$ is the set of negative samples and $\bf{v}_O$ is the output vector of node $\rm v_O$ that corresponds to the column $\rm O$ of matrix $\bf M$ in the skip-gram model. Then the training objective is simplified to be able to discriminate the output node $\rm v_O$ from nodes draws from $P_n(\rm v)$ using logistic regression. To update node embeddings under NEG, the derivative of loss $L$ with regard to the input of output $\rm v_j$ is given by
\begin{equation}\label{21}
\frac{{\partial L}}{{\partial {{\bf{v}}_{\rm{j}}}^{\rm{T}}\Phi ({{\rm{v}}_{\rm{I}}})}} = \sigma ({{\bf{v}}_{\rm{j}}}^{\rm{T}}\Phi ({{\rm{v}}_{\rm{I}}})) - {s_j},
\end{equation}
where $s_j$ is a binary indicator of samples, i.e., $s_j = 0$ if $\rm v_j$ is a negative sample, otherwise $s_j = 1$. 
The output vector is updated by 
\begin{equation}\label{22}
{{\bf{v}}_{\rm{j}}} \leftarrow {{\bf{v}}_{\rm{j}}} - \eta \left( {\sigma ({{\bf{v}}_{\rm{j}}}^{\rm{T}}\Phi ({{\rm{v}}_{\rm{I}}})) - {s_j}} \right)\Phi ({{\rm{v}}_{\rm{I}}})
\end{equation}
By using NEG, we just need to update the output vectors of nodes from ${\{ {{\rm{v}}_{\rm{O}}}\}  \cup {\rm{V}}_{neg}}$ instead of the entire node set $V$. The computational effort is therefore saved significantly.
Finally, the node vector is updated accordingly by the error backpropagation to the hidden layer, i.e.,
\begin{equation}\label{23}
\Phi ({{\rm{v}}_{\rm{I}}}) \leftarrow \Phi ({{\rm{v}}_{\rm{I}}}) - \eta \sum\limits_{{{\rm{v}}_{\rm{O}}} \in context({{\rm{v}}_{\rm{I}}})} {\sum\limits_{{{\rm{v}}_{\rm{j}}} \in \{ {{\rm{v}}_{\rm{O}}}\}  \cup {\rm{V}}neg} {\left( {\sigma ({{\bf{v}}_{\rm{j}}}^{\rm{T}}\Phi ({{\rm{v}}_{\rm{I}}})) - {s_j}} \right){{\bf{v}}_{\rm{j}}}} } 
\end{equation}

\begin{table}[tbp]
	\centering
	\tiny
	\setlength{\abovecaptionskip}{2pt}
	\caption{Summary of optimization techniques}
	\begin{tabular}{|p{1.5cm}<{\centering}|p{1.5cm}<{\centering}|p{3.5cm}<{\centering}|p{3.5cm}<{\centering}|l|}
		\hline
		\textbf{Technique} & \textbf{Working stage} & \textbf{Technical principle} & \textbf{Optimization goal} & \textbf{Publication} \\
		\hline
		Hierarchical Softmax & Node embedding & Leverage a tree structure to minimize the computation of conditional probabilities & Computational complexity reduction, acceleration of embedding process & \cite{DBLP:conf/aistats/MorinB05,DBLP:conf/kdd/PerozziAS14} \\
		\hline
		Negative Sampling & Feature extraction & Reduce the number of output vectors that need to be updated &  Computational complexity reduction, acceleration of embedding process & \cite{DBLP:journals/jmlr/GutmannH12,DBLP:conf/nips/MikolovSCCD13} \\
		\hline
		Attention Mechanism & (1) Feature extraction; (2) Node embedding & (1) Replace previously fixed hyperparameters with trainable ones; (2) Distinguish neighborhood's importance via trainable weights  &  Enhancement of training efficiency and effectiveness  & \cite{DBLP:conf/nips/Abu-El-HaijaPAA18,DBLP:conf/iclr/VelickovicCCRLB18} \\
		\hline
   \end{tabular}%
	\label{tab5}%
\end{table}%

\textbf{Attention Mechanism} 
Ever since attention mechanism was proposed, it has become an effective way to help models focus on the most important part of data. NRL also benefits from attention mechanism by conducting attention-guided random walks in feature extraction stage, and designing attention-based encoder 
in node embedding stage. 

In the stage of feature extraction, the attention mechanism \cite{DBLP:conf/nips/Abu-El-HaijaPAA18} is borrowed to lead  random walk to optimize an upstream objective. Let ${\bf{\Gamma }}$ be the transition matrix, ${{{\bf{\tilde P}}}^{(0)}}$ be the initial positions matrix with ${{{\bf{\tilde P}}}_{\rm vv}^{(0)}}$ set to the number of walks starting at node $\rm v$, and $C$ be the walk length, the context distribution can be represented by a $C$-dimensional vector $\bf{Q}=(Q_1,Q_2,...,Q_C)$. To obtain the expectation on co-occurrence matrix, $\mathbb{E}[\bf D]$, $\bf{Q}_k$ need to be assigned to $\bf{\Gamma}^k$ as a co-efficient. An attention model is proposed to learn $\bf Q$ automatically, where $\bf{Q}=\rm softmax((q_1,q_2,...,q_C))$, and all $\rm{q}_k$ can be trained by backpropagation. The attention model aims to guide the random surfer on "where to attend to" as a function of distance. To this end, the model on $\bf{\Gamma}^\infty$ is trained according to the expectation on random walk matrix, i.e.,
\begin{equation}\label{24}
\mathbb{E}[{{\bf{D}}^{{\rm{softmax[}}\infty ]}};{{\rm{q}}_{\rm{1}}},{{\rm{q}}_{\rm{2}}},...,{{\rm{q}}_\infty }] = {{{\bf{\tilde P}}}^{(0)}}\mathop {\lim }\limits_{C \to \infty } \sum\limits_{\rm k = 1}^C {{\rm{softmax(}}{{\rm{q}}_{\rm{1}}},{{\rm{q}}_{\rm{2}}},...,{{\rm{q}}_k}{\rm{)}}{\bf {\Gamma}^{\rm k}}} .
\end{equation}
For many random walk based methods like DeepWalk, they are special cases of the above equation where $C$ is not infinite, $\bf Q$ are fixed apriori, i.e., ${{\bf{Q}}_{\rm{k}}} = \left[ {1 - \frac{{{\rm{k}} - 1}}{C}} \right]$.

In the stage of node embedding, the graph attention network (GAT) \cite{DBLP:conf/iclr/VelickovicCCRLB18} proposes to  incorporate attention mechanism into a spatial GCN for providing differentiated weights of neighborhoods. Specifically, GAT defines a graph attention layer parametrized by a weight vector $\bf a$ and builds a graph attention network by stacking the layers. The attention function $\alpha(\cdot)$ is defined to measure the importance of neighbor $\rm u$ to the central node $\rm v$,
\begin{equation}\label{25}
{\alpha _{{\rm{vu}}}} = \frac{{\exp ({\rm{LeakyReLU(}}{{\bf{a}}^{\rm{T}}}{\rm{[}}{\bf{W}}{{\bf{h}}_{\rm{v}}}{\rm{||}}{\bf{W}}{{\bf{h}}_{\rm{u}}}{\rm{])}})}}{{\sum\nolimits_{{\rm{w}} \in {{\rm{N}}_{\rm{v}}}} {\exp ({\rm{LeakyReLU(}}{{\bf{a}}^{\rm{T}}}{\rm{[}}{\bf{W}}{{\bf{h}}_{\rm{v}}}{\rm{||}}{\bf{W}}{{\bf{h}}_{\rm{w}}}{\rm{])}})} }},
\end{equation}
where $\bf{h}_v$ refers to the feature of input node and $\bf W$ denotes the weight matrix of a shared linear transformation which is  applied to every node. The convolution operation (aggregation encoder) is defined as
\begin{equation}\label{26}
{\bf{h}}_{\rm{v}}^{\rm{t}} = \sigma \left( {\sum\limits_{{\rm{u}} \in {{\rm{N}}_{\rm{v}}}} {\alpha ({\bf{h}}_{\rm{v}}^{{\rm{t - 1}}},{\bf{h}}_{\rm{u}}^{{\rm{t - 1}}}){{\bf{W}}^{{\rm{t - 1}}}}{\bf{h}}_{\rm{u}}^{{\rm{t - 1}}}} } \right).
\end{equation}

\section{Recent Advances in Network Representation Learning}
Here we review existing NRL studies with a focus on recent methods that have achieved significant advances in machine learning and/or data mining. These studies are classified into four categories and 
the detailed taxonomy is shown in Fig. 4 which enables us to classify current NRL methods based on whether the learning is: 1) performed in centralized or distributed fashion; 2) using single network or multiple networks; 3) applied to static network or dynamic evolving network. Besides, we also include some works on knowledge graph as one of important extensions of NRL.
We review these works according to the proposed reference framework and present a brief summary as shown in Table 6.
\begin{figure}[t]
	\centering
	\includegraphics[width=0.85 \linewidth]{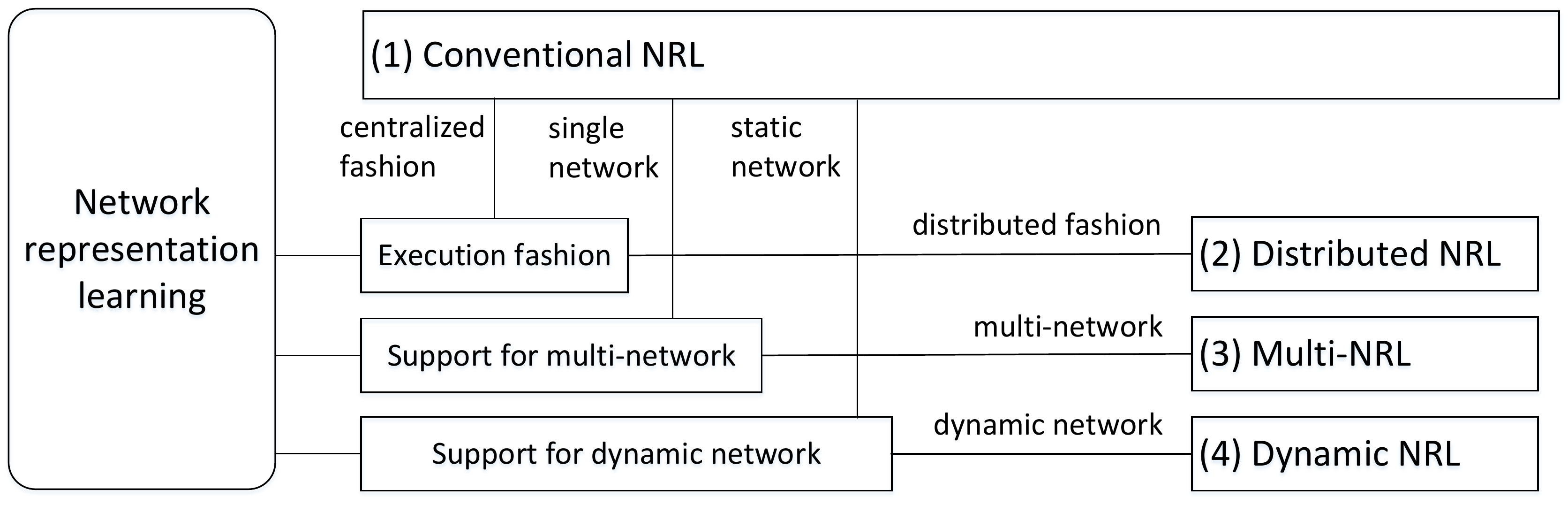}
	\caption{A taxonomy of NRLs based on its evolution directions.}
	\Description{NRL category}
\end{figure}
\subsection{Conventional NRL Methods}
Methods in this category share some common features such as performing in a centralized fashion, applying to only single network representation learning and only learning from a static network. We introduce recent advances in this category according to two main types of network properties, i.e., pairwise proximity and community structure.
\subsubsection{Pairwise Proximity}
MVE \cite{DBLP:conf/cikm/QuTSR0017} presents a multi-view embedding framework where pairs of nodes with different views are sampled as instances. Each view corresponds to a type of proximity between nodes, for example, the following-followee, reply, retweet relationships in many social networks. These views are usually complementary to each other. MVE uses skip-gram model to yield the view-specific representations that preserve the first-order proximities encoded in different views. Moreover, an attention-based voting scheme is proposed to identify important views by learning different weights of views.

SDNE \cite{DBLP:conf/kdd/WangC016} presents a semi-supervised learning model to learn node representations by capturing both local and global structure. The model architecture is illustrated in Fig. 5 that consists of two components: unsupervised component and supervised component. The former is designed to learn node representations by preserving the second-order proximity and the latter utilizes node connections as the supervised information to exploit the first-order proximity and refine node representations.
Specifically, given the adjacency matrix ${\bf{A}}$, the first component utilizes a deep autoencoder to reconstruct the neighborhood structure of each node. The encoder relies on multiple non-linear functions to encode each node $\rm v_i$ into a vector representation. The hidden representation in the $k$th layer is defined as
\begin{equation}\label{33}
{\bf{y}}_i^{(k)} = \sigma ({{\bf{W}}^{(k)}}{\bf{y}}_i^{(k - 1)} + {{\bf{b}}^{(k)}}),
\end{equation}
where $\bf x_i$ is the input vector of $\rm v_i$, ${\bf{y}}_i^{(0)} = {{\bf{x}}_i}$, $\bf W^{(k)}$ and $\bf b^{(k)}$ are weights and biases respectively in the $k$th layer. The decoder reconstructs the input vectors (e.g., ${\bf{\hat x}}_i$) from the most hidden vectors (e.g., ${\bf{y}}_i^{(k)}$) by means of non-linear functions. Note that the number of zero elements in  ${\bf{A}}$ is far less than that of zero elements. The autoencoder is prone to reconstruct the zero elements. To avoid this situation, SDNE imposes more penalty to the reconstruction error of non-zero elements and the loss function is defined as
\begin{equation}\label{34}
{L_{2nd}} = \sum\limits_{i = 1}^n {\left\| {({{{\bf{\hat x}}}_i} - {{\bf{x}}_i}) \circ {{\bf{b}}_i}} \right\|_2^2},
\end{equation}
where $\circ$ denotes the Hadamard product. The second component enhances the first-order proximity by borrowing the idea of Laplacian Eigenmaps to incur a penalty once neighboring nodes are embedded far away.
Consequently, the loss function is defined as
\begin{equation}\label{35}
{L_{1st}} = \sum\limits_{i,j = 1}^n {{{\bf{A}}_{ij}}\left\| {({\bf{y}}_i^{(k)} - {\bf{y}}_j^{(k)}) \circ {{\bf{b}}_i}} \right\|_2^2}. 
\end{equation}
In addition, in order to avoid falling to local optima in the parameter space, SDNE leverages deep belief network to pretrain the parameters at first.

\begin{figure}[t]
	\centering
	\includegraphics[width=.7\linewidth]{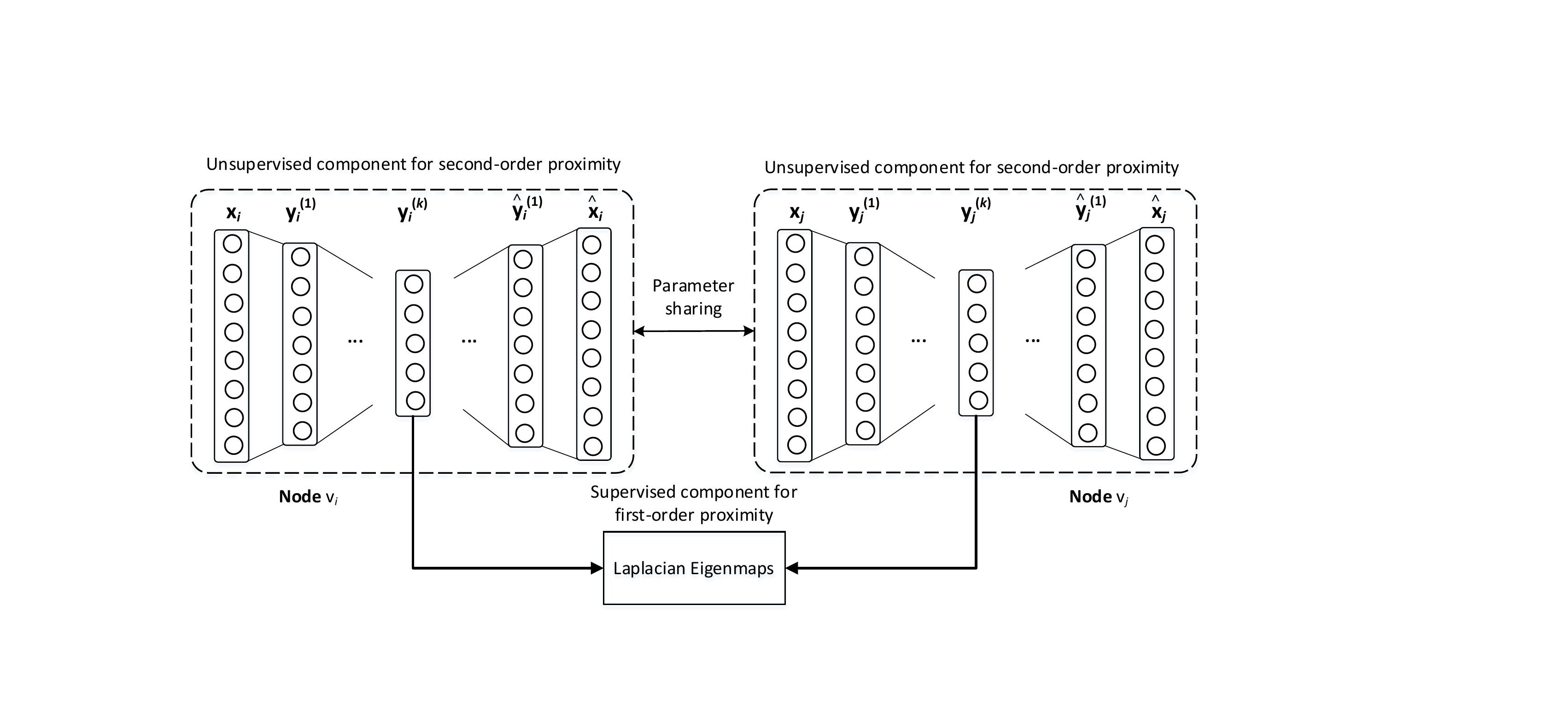}
	\caption{The semi-supervised model of SDNE (adapted from Fig. 2 \cite{DBLP:conf/kdd/WangC016}).}
	\Description{SDNE}
\end{figure}

DNGR \cite{DBLP:conf/aaai/CaoLX16} designs a deep denoising autoencoder method to capture the non-linearities of network features. Its basic idea is to learn node representations from the positive pointwise mutual information (PPMI) matrix of nodes and their contexts. As illustrated in Fig. 6, DNGR consists of three components: random surfing, calculation of PPMI and a stacked denoising autoencoder (SDAE). In the first component, random surfing is used to generate the probabilistic co-occurrence (PCO) matrix that corresponds to a transition matrix by nature. Then the second component calculates the PPMI matrix based on the PCO matrix by following \cite{DBLP:journals/brm/Bullinaria07}. After that, SDAE is presented for highly non-linear abstractions learning. In order to recover the complete matrix under certain assumptions, SDAE partially corrupt the training sample $\bf x$ by randomly assigning some of $\bf x$'s entries to zero with a certain probability. As a result, the objective becomes minimizing the reconstruction loss, i.e.,
\begin{equation}\label{36}
\mathop {\min }\limits_{{\theta _1},{\theta _2}} \sum\limits_{i = 1}^n {L({{\bf{x}}^{(i)}},{g_{{\theta _2}}}({f_{{\theta _1}}}({{{\bf{\tilde x}}}^{(i)}})))},
\end{equation}
where ${{f_{{\theta _1}}}}(\cdot)$ denotes an encoding function, ${{g_{{\theta _2}}}}(\cdot)$ denotes a decoding function, the $i$th instance and its corrupted form are denoted by ${\bf x}^{(i)}$ and ${\bf{\tilde x}}^{(i)}$ respectively.

Struc2vec \cite{DBLP:conf/kdd/RibeiroSF17} insists that node identity similarity should be independent of network position and neighborhoods' labels. To well preserve the property, the input network $G$ is firstly preprocessed into a context graph $M$ which is a multi-layer weighted graph described in Sec. 3.1.1. Then a biased random walk process is conducted to produce node traveling sequences. Each walker chooses its next step on the same layer or across different layers and the choosing probabilities are proportional to edge weights, so that the structurally similar nodes are more likely to be visited. After having samples, Struc2vec uses skip-gram to train the learning model and hierarchical softmax is leveraged to minimize the complexity.
\subsubsection{Community Structure}
MRF \cite{DBLP:conf/aaai/JinYLHCFC19} proposes a structured pairwise Markov random field framework. To ensure the global coherent community structure, MRF adopts the Gibbs distribution to measure the posterior probability $P(C_p|\bf A)$ of community partition $C_p$ given network adjacency matrix $\bf A$ and maximize $P(C_p|\bf A)$ by optimizing a well-designed energy function $E(C_p;\bf A)$. $E(C_p;\bf A)$ consists of two parts: a group of unary potentials that are used to enforce node representations playing a dominant role and a group of pairwise potentials that are used to fine-tune the obtained unary potentials based on the connections of nodes. Besides, Gaussian mixture model is utilized to approximate the probability distributions of all nodes belonging to various communities.

To capture the overlapping community structure, Epasto et al. \cite{DBLP:conf/www/EpastoP19} propose a multi-embedding method, splitter, to learn multiple vectors for each node, representing each one's involvement in various communities. In order to exploit the multi-community participation, splitter preprocesses the original network into a persona graph $G_p$ (see Sec. 3.1.1) having multiple personas for every node. But the persona graph may consists of many disconnected components that pose challenges for representation learning. To address the challenge, splitter adds virtual edges between each persona and its parent node in the original network so as to enforce that every original node $\rm v_o$ can be predicted given its arbitrary persona $\rm v_i$'s representation, i.e., $\Pr ({{\rm{v}}_{\rm{o}}}|{\Phi _{{G_p}}}({{\rm{v}}_{\rm{i}}}))$. Like DeepWalk, splitter uses skip-gram model coupled with hierarchical softmax to learn node representations.
By introducing a regularization parameter $\lambda$, the optimization objective is formalized as follows:
\begin{equation}\label{32}
\mathop {\arg \min }\limits_{{\Phi _{{G_p}}}}  - \log \Pr (\{ {{\rm{v}}_{{\rm{i - w}}}},...,{{\rm{v}}_{{\rm{i + w}}}}\} \backslash {{\rm{v}}_{\rm{i}}}|{\Phi _{{G_p}}}({{\rm{v}}_{\rm{i}}})) - \lambda \Pr ({{\rm{v}}_{\rm{o}}}|{\Phi _{{G_p}}}({{\rm{v}}_{\rm{i}}})).
\end{equation}

\begin{figure}[t]
	\centering
	\includegraphics[width=1\linewidth]{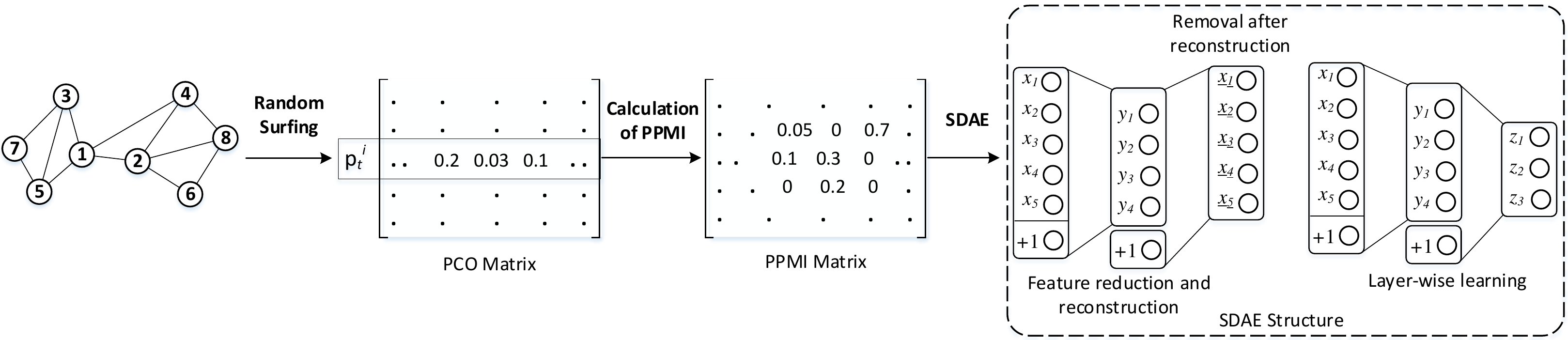}
	\caption{The framework of DNGR (adapted from Fig. 1 \cite{DBLP:conf/aaai/CaoLX16}).}
	\Description{DNGR}
\end{figure}
vGraph \cite{DBLP:conf/nips/SunQHH019} presents a probabilistic generative model where each node $\rm v$ is viewed as a mixture of multiple communities and represented by a multinomial distribution over communities $C$, i.e., $p(C|\rm v)$, meanwhile each community $C$ can be represented by a distribution over nodes, i.e., $p(\rm v|$$C)$. vGraph casts the edge generation process as an inference problem. Specifically, for each node $\rm v$, vGraph first draws a community assignment $C$ from $p(C|\rm v)$ and then generates an edge $\rm e_{vu}$ by drawing another node $\rm u$ according to distribution $p(u|\rm C)$. Both types of distributions are parameterized by the representations of nodes and communities. Besides, a smoothness regularizer is borrowed by vGraph to ensure the community memberships of neighborhoods to be similar.

\subsection{Distributed NRL Methods}
Considering the recent advances in GPU-enabled neural network training, Zhu et al. \cite{DBLP:conf/www/ZhuXTQ19} propose a CPU-GPU hybrid system, GraphVite, which is a hardware coupling system specifically designed for large-scale network embedding. To leverage distinct advantages of CPUs and GPUs, GraphVite focuses on the parallelization of instance sampling inside of feature extraction and node embedding, and its overview is shown in Fig. 7. The random walk based sampling procedure involves excessive random access and the random walks initiated from different starting nodes are independent, so the positive instance sampling is suitable for parallel execution on multiple CPUs. While the node embedding procedure involves excessive matrix computation which is the advantage of GPUs. 
Due to the limited GPU memory, it is impossible to place all sampling instances and parameter matrices of node embeddings on a GPU, GraphVite organizes a grid sample pool. The pool is divided into multiple blocks and each one corresponds to a subset of the original network. The training task is also divided into many subtasks and they are assigned with varied blocks for training. Due to the sparse nature of networks, many block pairs are gradient exchangeable, which means exchanging the order of gradient descent steps does not result in a vector difference. As a result, subtasks on different GPUs are capable of performing gradient updates currently in their own subsets without any synchronization. At the same time, the sampling pool is shared between CPUs and GPUs. In order to reduce the synchronization cost, GraphVite presents a collaboration strategy which maintains two sample pools in the main memory, so that CPUs fill up a pool and pass it to GPUs and they always work on different pools in a parallel fashion. GraphVite reports about 50 times faster than the current fastest system and takes around 20 hours to embed a network as large as 66 million nodes and 1.8 billion edges.
\begin{figure}[t]
	\centering
	\includegraphics[width=0.95 \linewidth]{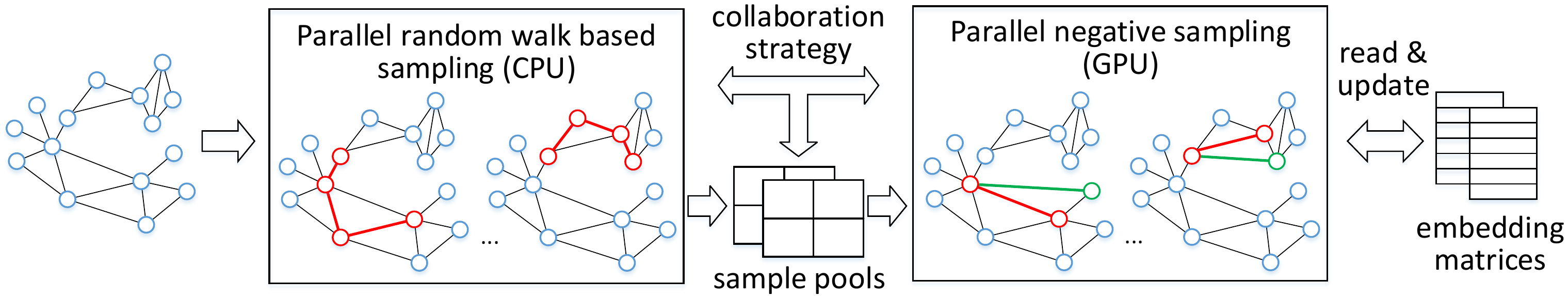}
	\caption{Overview of the CPU-GPU hybrid system (adapted from Fig. 1 \cite{DBLP:conf/www/ZhuXTQ19}).}
	\Description{Graphvite}
\end{figure}
\begin{figure}[t]
	\centering
	\includegraphics[width=0.75 \linewidth]{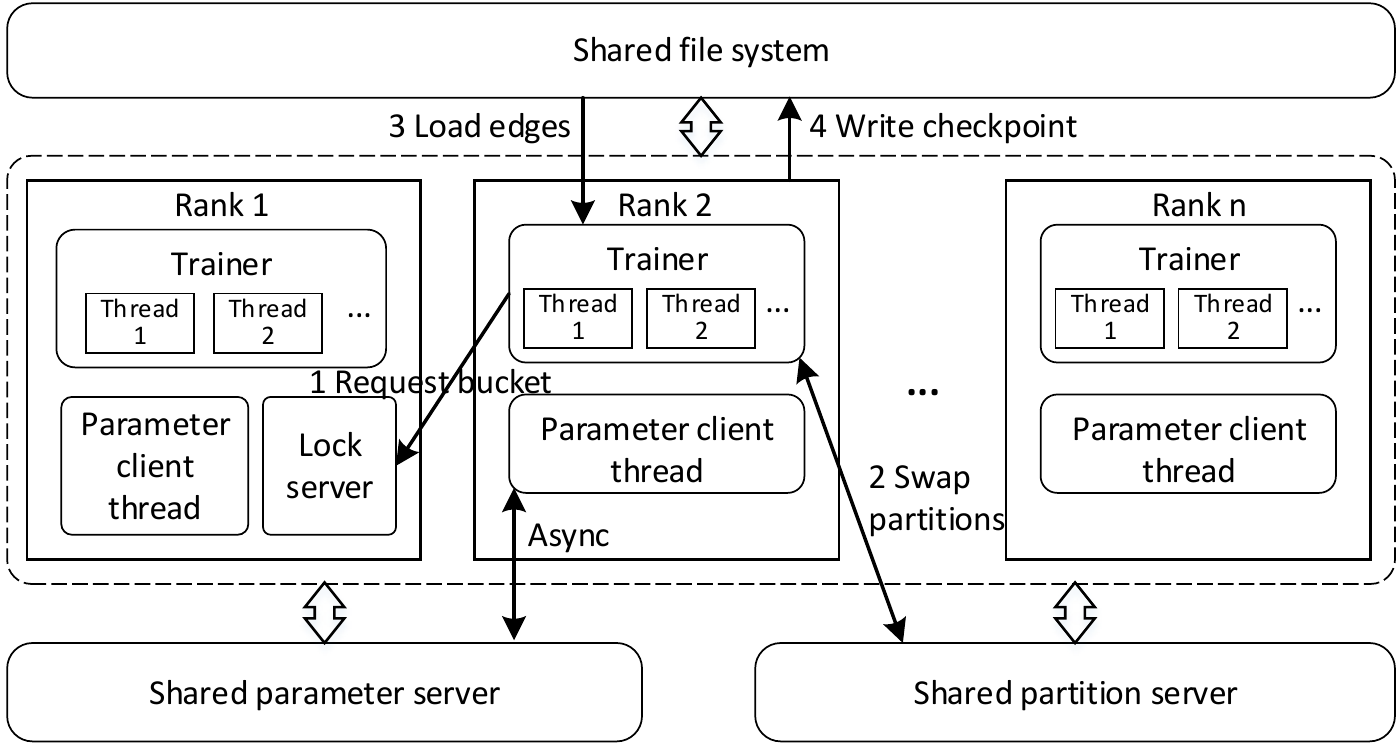}
	\caption{A block diagram of the modules used for PBG's distributed mode (adapted from Fig. 2 \cite{DBLP:sysml/abs-1903-12287}).}
	\Description{PBG}
\end{figure}

Lerer et al. \cite{DBLP:sysml/abs-1903-12287} also present a distributed embedding system, PyTorch-BigGraph (PBG), which incorporates several modifications to traditional embedding systems so as to allow it supports scale to multi-entity, multi-relation graphs with billions of nodes and trillions of edges. To implement parallelization, PBG partitions the adjacency matrix of original graph into multiple buckets and feeds the edge samples from buckets distributively. An example of the training of one bucket upon rank 2 is illustrated in Fig. 8. The trainer requests a bucket from the lock server residing on rank 1, where all partitions of that bucket are locked. Then the trainer saves the no longer used partitions and loads new partitions that it needs to and from the shared partition servers, at which point it drops its old partitions left on the lock server. Edge samples are loaded from a shared filesystem and the training is performed inside multi-thread with no inter-thread synchronization required. In a single thread, only a small fraction of shared parameters will be synchronized with a shared parameter server. Note that checkpoints will be occasionally written back to the filesystem from trainers. A distributed execution of PBG on 8 machines achieves 4x speedup and reduces memory consumption by 88\%.


\subsection{Multi-NRL Methods}
Compared to single network embedding, multiple networks may contain complementary information and multi-network embedding can produce better representations. Similar examples can be observed in many fields like social networks, bioinformatics, etc. a node in one network may be associated with multiple nodes in another network which forms a many-to-many mapping relationship between networks. Ni et al. \cite{DBLP:conf/www/NiCLCCX018} propose a deep multi-network embedding (DMNE) method, which coordinates different neural networks with one co-regularized loss to manipulate cross-network correlations. An example of DMNE for two networks is illustrated in Fig. 9, where $\bf{A}^{(1)}$ and $\bf{A}^{(2)}$ are network contexts derived from the two networks, $\bf{S}^{(12)}$ and $\bf{S}^{(21)}$ denote the cross-network relationship matrix, and $(\bf{H}^{(i)})_l$ corresponds to the matrix of representations for all nodes in the $l$th layer of network $i$. For single network embedding say network $i$, the neural network consists of $L_i + 1$ layers built in the autoencoder fashion, where half of hidden layers act as encoders to learn node representations, while the others are decoders who are in charge of the reconstruction of input. To achieve the network information complementarity, DMNE relies on the intuition that a node's representation should be similar to the representation of its mapped node in another network, and introduces the cross-network regularization based on two kinds of loss functions: embedding disagreement (ED) loss function and proximity disagreement (PD) loss function. The former is for situation that all networks have an identical embedding dimension, while the latter is more flexible without dimension constraint. As a result, the unified objective is represented by
\begin{equation}\label{key33}
\mathop {\arg \min }\limits_{{{\bf{\theta }}^{(i)}},{{\bf{U}}^{(i)}}} \sum\limits_{i = 1}^g {L_{ae}^{(i)}}  + \alpha \sum\limits_{(i,j) \in I}^g {L_R^{(i)}}  + \beta \sum\limits_{i = 1}^g {\left\| {{{\bf{U}}^{(i)}} - {{\bf{H}}^{(i)}}} \right\|_F^2} 
\end{equation}
where $\bf{U}^{(i)}$ denotes network $i$'s representation matrix, $\bf{\theta}^{(i)}$ denotes the weight matrix, $L_{ae}^{(i)}$ refers to the loss of network $i$, $L_{R}^{(ij)}$ can be either ED loss or PD loss, and $\alpha$ and $\beta$ are trade-off factors.

TransLink \cite{DBLP:conf/infocom/ZhouF19} models social network alignment as a link prediction between different networks and aims to find out all pairs of user accounts with same identities and connects them via anchor links. TransLink embeds users and links between uses into a latent space by incorporating both network structure and user interaction meta-path, and then iteratively predict the potential anchor links between users who have similar representations over translative operations.
Du et al. \cite{DBLP:conf/ijcai/DuYZ19} investigate a joint framework, CENALP that couples link prediction and network alignment together. The newly predicted links enrich the network structure information, and new potential nodes with richer features will have a greater chance to be identified and vice versa. Therefore, link prediction and network alignment are allowed to work in a mutually beneficial way. CENALP relies on a biased random walk to generate samples across different networks. For each training sample ($\rm v_i, v_j$), CENALP adopts a product layer to decode the latent interaction between two nodes and sets up a logistic regression layer for prediction.
\begin{figure}[t]
	\centering
	\includegraphics[width=0.95 \linewidth]{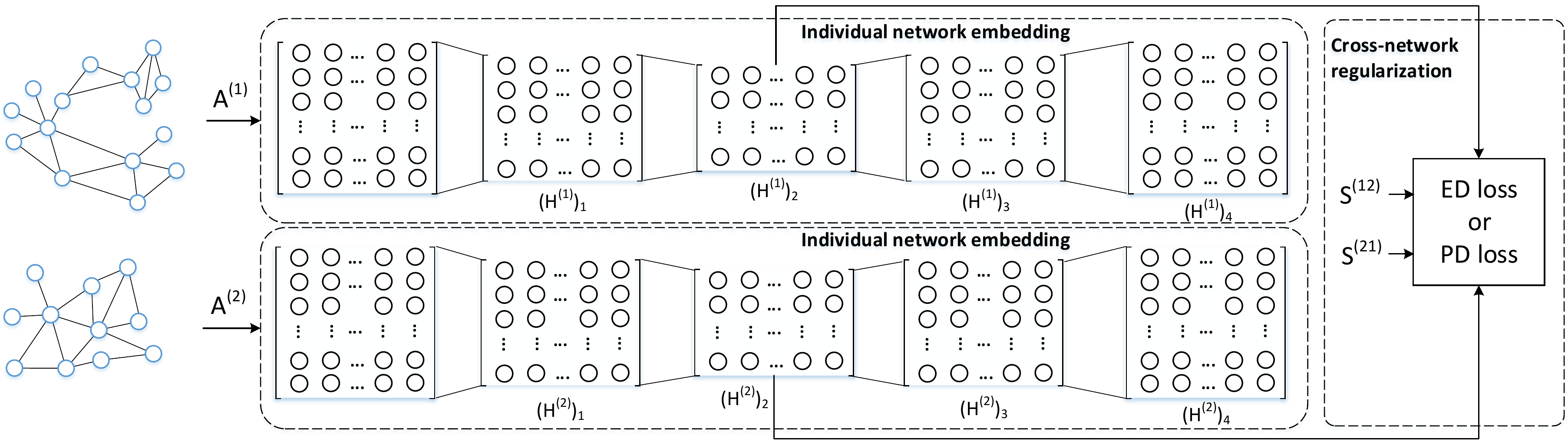}
	\caption{The DMNE framework for two networks (adapted from Fig. 2 \cite{DBLP:conf/www/NiCLCCX018}).}
	\Description{DMNE}
\end{figure}

\subsection{Dynamic NRL Methods}
The methods introduced so far mainly focus on exploring NRL for static networks. However, real-world networks like social networks and biological networks are dynamically evolving over time, which means not all nodes are available during the training. Inductive learning is an effective way to support the representation learning for unseen nodes. As the first inductive framework, GraphSage \cite{DBLP:conf/nips/HamiltonYL17} derives from the idea of Spatial GCN where in order to generate node embeddings.
It designs an aggregation function to essentially assemble features from each node's neighborhood, e.g., text attributes, node degrees. The aggregation encoder is defined as follows:
\begin{equation}\label{37}
{\bf{h}}_{\rm{v}}^{\rm{k}} = \sigma ({\bf{W}^k} \cdot AGG_k({\bf{h}}_{\rm{v}}^{{\rm{k - 1}}},\{ {\bf{h}}_{\rm{u}}^{{\rm{k - 1}}},\forall {\rm{u}} \in {N_{\rm{v}}}\} )),
\end{equation} 
where the function $AGG_k(\cdot,\cdot)$ should be symmetric so as to ensure the framework can be trained and applied to neighborhood feature sets with arbitrary order. GraphSage examines three types of functions, i.e., mean aggregator, LSTM aggregator and pooling aggregator. Note that LSTM is not inherently permutation invariant, so it is applied to a random permutation of neighborhoods for implementation. The learning procedure of GraphSage is illustrated by an example in Fig. 10. For each node to be embedded, GraphSage first samples a fixed number of neighborhoods within $k$ hops, then obtains the central node's state by aggregating its neighborhoods' features, and finally makes predictions and backpropagate errors based on the node's state. In order to learn useful predictive representations, negative sampling is introduced to enhance performance, and then a graph-based loss function is given by
\begin{equation}\label{key35}
L({{\bf{z}}_{\rm{u}}}) =  - \log (\sigma ({\bf{z}}_{\rm{u}}^{\rm{T}}{{\bf{z}}_{\rm{v}}})) - Q \cdot {E_{{{\rm{v}}_{\rm{n}}} \sim {P_{n({\rm{v}})}}}}\left[ {\log \sigma ( - {\bf{z}}_{\rm{u}}^{\rm{T}}{{\bf{z}}_{\rm{n}}})} \right],
\end{equation}
where $\bf z_u$ denotes the representation of node $\rm v_u$, $Q$ denotes the amount of negative samples and $P_n$ represents the negative sampling distribution. As an inductive representation learning method, GraphSage enforces that every representation $\rm v_u$ fed into the above function is generated from the features of node $\rm u$'s neighborhoods rather than a unique representation trained by an embedding look-up.
\begin{figure}[t]
	\centering
	\includegraphics[width=0.8 \linewidth]{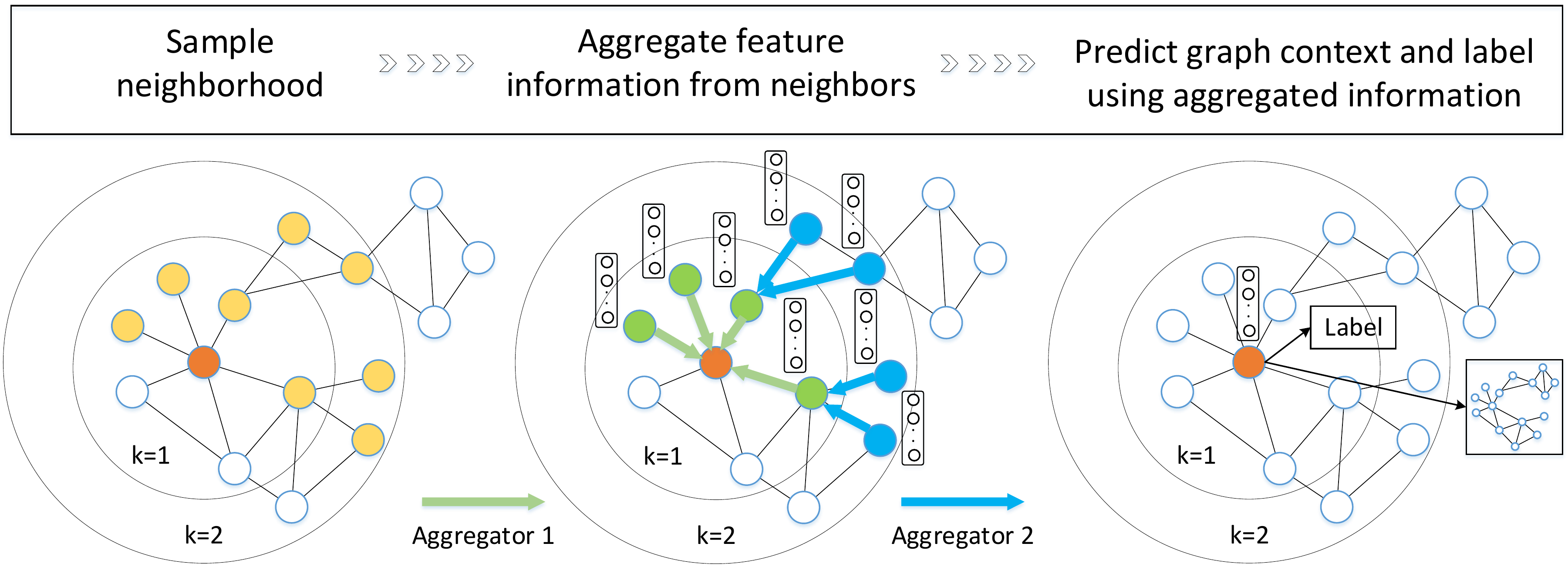}
	\caption{An example illustrating the three-step learning procedure of GraphSage (adapted from Fig. 1 \cite{DBLP:conf/nips/HamiltonYL17}).}
	\Description{GraphSage introduction}
\end{figure}

Graph2Gauss (G2G) \cite{DBLP:conf/iclr/BojchevskiG18} proposes a novel unsupervised inductive learning method that embeds nodes in (un)directed attributed graphs into Gaussian distributions rather than conventional vector representations, so that it can capture the uncertainty of each node's embedding. G2G first uses a deep encoder to yield the parameters associated with the node's embedding distribution from the node attributes. 
The mean and diagonal representations for a node are learned as functions of the node attributes. After that, G2G optimizes a ranking loss function that incorporates the ranking of nodes derived from the network structure. Given an anchor node $\rm v$, nodes at distance 1 of $\rm v$ are closer than nodes at distance 2, etc. Distance between node representations is measured by the asymmetric KL divergence. The ranking loss is a square-exponential loss proposed in energy based models. 
To avoid the situation that low-degree nodes are less often updated, G2G presents a node-anchored sampling method. For each node $\rm v$, the method randomly samples one other node from its neighborhoods and then optimize over all the corresponding pairwise constraints. To enable the inductive learning for unseen nodes, G2G passes the attributes of these unseen nodes through the learned deep encoder. 

To incorporate global structural information, SPINE \cite{DBLP:conf/ijcai/GuoXL19} proposes a structural identity preserved method. Rooted PageRank matrix $\bf{S}^{RPR}$ is used as the indicator of pairwise proximity where $\bf{S}_i^{RPR}$ represents $\rm v_i$'s global feature. 
As an inductive method, the length of a node's structural description should be independent of the network size, so that the structural feature $\bf {T}_i$ of $\rm v_i$ is defined as the top-$k$ values of $\bf{S}_i^{RPR}$. Given the content feature matrix $\bf{F}_i^k$ of $k$ nodes, node $\rm v_i$'s embedding is generated by aggregating the $k$ vectors w.r.t. the corresponding weights in $\bf T_i$. To encode the similarity in terms of both structural identities and local proximities simultaneously, a biased positive sampling strategy is proposed to collaborate with negative sampling.

Zhou et al. \cite{DBLP:conf/aaai/ZhouYR0Z18} proposes a dynamic NRL method, DynamicTriad, to preserve both structural information and evolution patterns during the learning process. The evolution of a dynamic network is described by a series of static network snapshots over discrete time. A triad of three nodes works as the basic units of networks and its closure process is used to capture the network dynamics. In specific, the triad closure process reflects how a closed triad develops from an open triad over time by means of an evolutionary probability, where a closed triad is a complete graph of three nodes while an open triad misses a connection between any two nodes. For sampling, considering the expensive computation on the combination of positive and negative samples, DynamicTriad adopts an idea of sample corruption to generate negative samples by replacing nodes in positive triad with varied nodes.
Besides, DynamicTriad enforces the evolutionary smoothness by minimizing the distance of node representations in adjacent timestamps. In the dynamic settings, nodes not seen at the current timestamp are regarded as out-of-sample nodes and their embeddings can be inferred by exploring the idea of inductive learning. To ensure the inferred embeddings preserve intricate network properties, DepthLGP \cite{DBLP:conf/aaai/MaC018} designs a high-order Laplacian Gaussian process (hLGP) to encode these properties, and employs a DNN to learn a nonlinear transformation from the hLGP.
\begin{figure}[t]
	\centering
	\includegraphics[width=0.8 \linewidth]{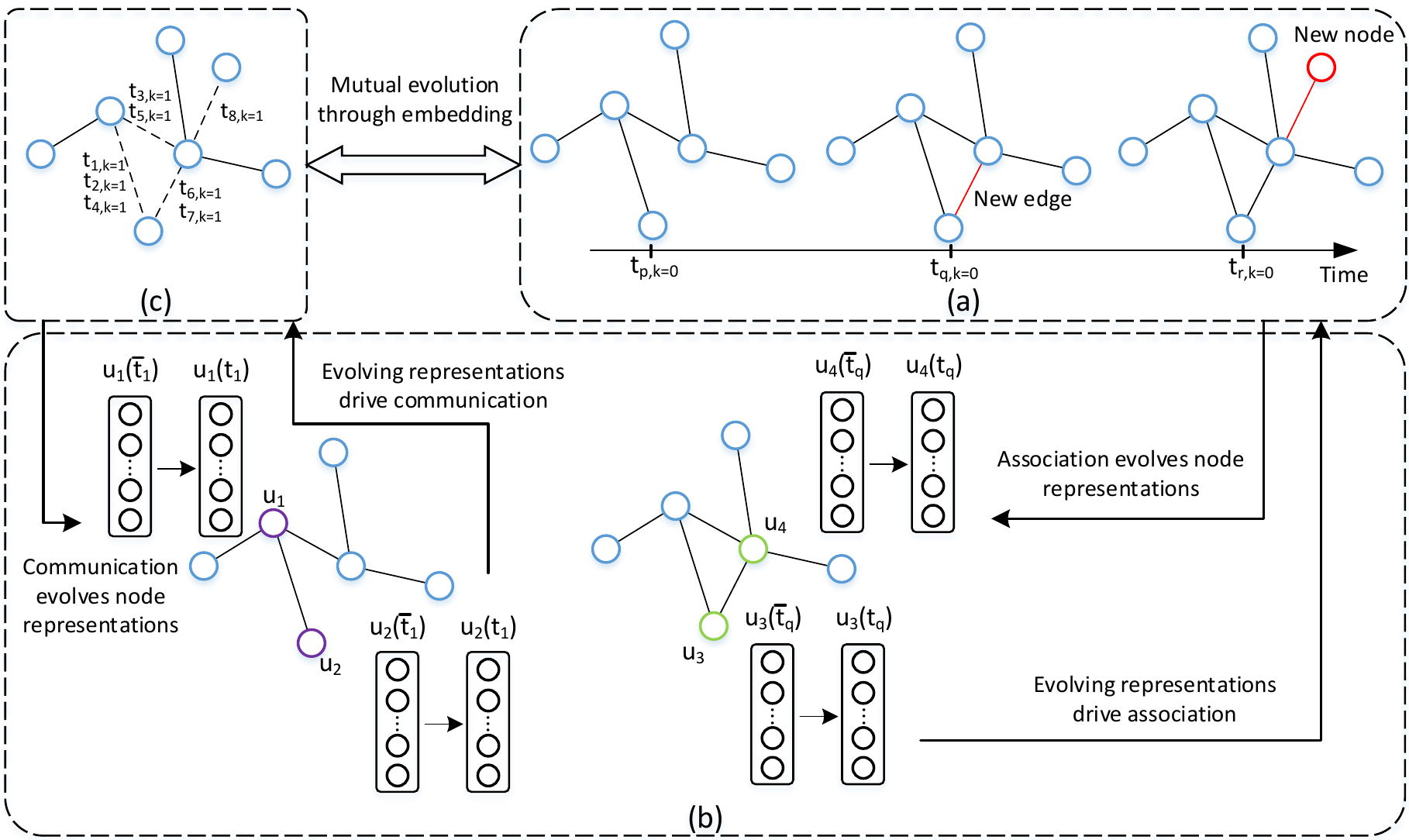}
	\caption{An illustration of evolution for representation learning. (a) Association events (k=0); (b) Evolving representations; (c) Communication events (k=1). (adapted from Fig. 1 \cite{DBLP:conf/iclr/TrivediFBZ19})}
	\Description{DyRep}
\end{figure}

The above work assume that network dynamics evolve at a single time scale, however, the evolution of real-world networks usually exhibit multiple time scales. DyRep \cite{DBLP:conf/iclr/TrivediFBZ19} uses two distinct dynamic processes to model the dynamic evolution: association process describes dynamics of the network, which brings structural changes caused by nodes and edges and results in long-lasting information flow associated with them; and communication process reflects dynamics on the network, which relates to the activities between connected or non-connected nodes and results in temporary information exchange across them. As shown in Fig. 11, the dynamic evolution of a network is given by a stream of events that involve both processes. To model the occurrence of event $p=(\rm u, v, t_p, k)$ between $\rm u$ and $\rm v$ at time $\rm t_p$, a conditional intensity function is defined based on the temporal point process, i.e.,
\begin{equation}\label{42}
\lambda _k^{{\rm{u,v}}}(t_p) = {f_k}(g_k^{{\rm{u,v}}}(\bar t_p)),
\end{equation}
where $\bar t_p$ is the timestamp before the current event, and $f_k(\cdot)$ is a parameterized softplus function designed for capturing the timescale dependence. Once an event $p$ occurred, the representation of each involved node will be updated via a deep recurrent neural network (RNN), e.g., node $\rm v$'s representation $\bf{z}^{v}(t_p)$ at current timestamp $t_p$ is updated by
\begin{equation}\label{43}
{{\bf{z}}^{\rm{v}}}({t_p}) = \sigma (\underbrace {{{\bf{W}}^{struct}}{\bf{h}}_{struct}^{\rm{u}}({{\bar t}_p})}_{{\rm{Localized \, embedding \, propagation}}} + \underbrace {{{\bf{W}}^{rec}}{{\bf{z}}^{\rm{v}}}(\bar t_p^{\rm{v}})}_{{\rm{Self - propagation}}} + \underbrace {{{\bf{W}}^t}({t_p} - \bar t_p^{\rm{v}})}_{{\rm{Exogenous \,  drive}}}),
\end{equation}
where ${\bf{h}}_{struct}^{\rm{u}}$ is the representation aggregated from node $\rm u$'s neighborhoods,  ${{{\bf{z}}^{\rm{v}}}(\bar t_p^{\rm{v}})}$ is the recurrent state obtained from $\rm v$'s previous representation, and ${\bar t_p^{\rm{v}}}$ denotes the timestamp of $\rm v$'s previous event. Parameter matrices ${{{\bf{W}}^{struct}}}$, ${{{\bf{W}}^{rec}}}$ and ${{{\bf{W}}^t}}$ are applied to control the aggregate effect of three inputs. For a set of observed events, the learning objective is defined as negative log likelihood. Specifically, DyRep borrows the idea of graph attention networks (GAT) \cite{DBLP:conf/iclr/VelickovicCCRLB18} that is the attention mechanism applied to graph data, and uses it to endue neighborhoods with varied attention coefficients that also evolve over time.
	\begin{figure}[t]
	\centering
	\includegraphics[width=0.99 \linewidth]{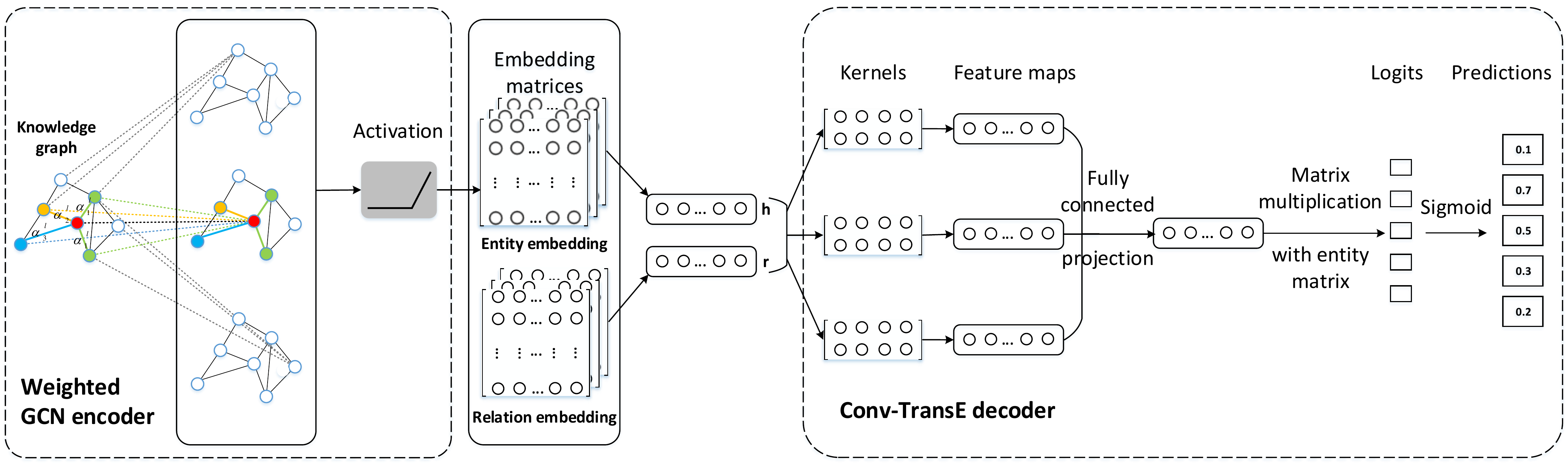}
	\caption{An illustration of SACN (adapted from Fig. 1 \cite{DBLP:conf/aaai/ShangTHBHZ19}).}
	\Description{SACN}
\end{figure}

\subsection{Knowledge Graph Representation Learning}
	As a special type of networks, knowledge graph can be viewed a structured representation of facts, denoted by a set of triples. Each triple consists of two entities h, t and a relation r between them, e.g., (h, r, t). Knowledge graph representation learning (KRL) is to map entity nodes and relation edges into low-dimensional vectors while capturing their semantic meanings \cite{DBLP:journals/corr/abs-2002-00388}. The primary goal of KRL is to improve the plausibility of facts where the plausibility can be viewed as a specialization of network property. 
	Distance proximity is often used to enhance the plausibility of a fact by minimizing the distance between entities of the corresponding triple. In the classic translating embedding model \cite{DBLP:conf/nips/BordesUGWY13}, the relation r corresponds to a translation from head entity to tail entity and their embeddings $\bf{h+r}\approx t$ hold when (h, r, t) is a fact. RotatE \cite{DBLP:conf/iclr/SunDNT19} expands the embedding space from real-valued space to complex space where entities and relations are mapped to low-dimensional complex vectors and each relation corresponds to a rotation from head entity to tail entity, i.e.,
	\begin{equation}\label{key38}
	\bf t\approx h \circ r,
	\end{equation} 
	where $\circ$ denotes the Hadamard product. For each element in the embedding space, $t_i \approx h_i \circ r_i$ and $|r_i|=1$. Element $r_i$ is of the form $e^{i\theta_{r,i} }$, which corresponds to a counterclockwise rotation by $\theta_{r,i}$ radians about the origin of the complex plane. For each triple (h, r, t), its distance is defined as
	\begin{equation}\label{key39}
	d_r(\bf{h,t})=\left\| {\bf{h \circ r}} - t \right\|.
	\end{equation}
	
	The complex space allows RotatE to capture more relation patterns including symmetry/antisymmetry, inversion and composition. During the training, traditional negative sampling samples the negative triples in a uniform way, which suffers the problem of inefficiency since many samples are obviously false and cannot provide meaningful information. RotatE presents a variant called self-adversarial negative sampling, which samples negative triples from the following distribution
	\begin{equation}\label{key40}
	p({h'_j},r,{t'_j}|\{ ({h_i},{r_i},{t_i})\} ) = \frac{{\exp \alpha {d_r}({\bf{h'}}_j, {\bf{t'}}_j)}}{{\sum\limits_i {\exp \alpha {d_r}({\bf{h'}}_i,{\bf{t'}}_i)} }},
	\end{equation}
	where $\alpha$ is the temperature of sampling. The above probability work as the weight of the negative sample. The negative sampling loss is defined as
	\begin{equation}\label{key41}
	L =  - \log \sigma (\gamma  - {d_r}({\bf{h,t}})) - \sum\limits_i {p({h'_i},r,{t'_i})\log \sigma ({d_r}({\bf{h'}}_i,{\bf{t'}}_i) - \gamma )} ,
	\end{equation}
	where $\gamma$ is a fixed margin, $\sigma$ is the sigmoid function, and $({h'_i},r,{t'_i})$ is the $i$th negative triple.

	SACN \cite{DBLP:conf/aaai/ShangTHBHZ19} proposes an end-to-end structure-aware convolutional network, SACN, to predict new triples for knowledge graph completion. As illustrated in Fig. 12, SACN presents a weighted GCN as the encoder to learn entities's representations by aggregating connected entities as specified by the relations in the knowledge graph, where the weighted GCN weights the different types of relations differently by defining the interaction strength, e.g., $\alpha_t$ denotes the strength of relation type $t$, so that the amount of information from neighboring nodes used in aggregation can be controlled. Let ${\bf{h}}_i^l$ be the input vector of node $\rm{v}_i$ in the $l$th layer, its output vector is given by 
	\begin{equation}\label{key42}
	{\bf{h}}_i^{l+1} = \sigma \left( {\sum\limits_{{\rm{v}}_j \in {N_i}} {\alpha _t^lg({\bf{h}}_i^{l},{\bf{h}}_j^{l})} } \right),
	\end{equation}
	where $N_i$ is the neighbor set of node ${\rm{v}}_i$ and $g$ is the aggregation function. With node embeddings as the input, the decoder aims to represent the relations more accurately by recovering the original triples. Based on ConvE \cite{DBLP:conf/aaai/DettmersMS018}, SACN develops Conv-TransE as the decoder where the translation fashion of TransE is incorporated. Conv-TransE aligns the convolutional outputs of both entity and relation embeddings with all kernels and yields a matrix $\bf{M}(h,r)$. Both encoder and decoder are jointly trained by minimizing the cross-entropy between $\bf{h+r}$ and $\bf t$ to preserve the distance proximity and the objective function is defined as
	\begin{equation}\label{key43}
	d_r({\bf{h,t}})=f(vec({\bf{M}(h,r)})\bf{W})\bf{t},
	\end{equation}
	where $f$ is a non-linear function and $W$ is a matrix for the linear transformation. 
	$vec(\bf M)$ corresponds to a reshaping operation that changes feature map matrix to a vector.
\begin{table}[htbp]
	\centering
	\tiny
	\setlength{\abovecaptionskip}{4pt}
	\caption{Summary of recent advances in NRL}
	\begin{tabular}{|p{1.1cm}<{\centering}|p{1.1cm}|p{1.8cm}<{\centering}|p{1.9cm}<{\centering}|p{1.5cm}<{\centering}|p{1.9cm}<{\centering}|p{1.9cm}<{\centering}|}
		\hline
		\multirow{2}[4]{*}{\textbf{Category}} & \multicolumn{1}{c|}{\multirow{2}[4]{*}{\textbf{NRL work}}} & \multicolumn{1}{c|}{\multirow{2}[4]{*}{\textbf{\tabincell{c}{Data preprocessing \\ method, input data}}}} & \multicolumn{2}{c|}{\textbf{Network feature extraction}} & \multicolumn{2}{c|}{\textbf{Node embedding}} \\
		\cline{4-7}          & \multicolumn{1}{c|}{} & \multicolumn{1}{c|}{} & \textbf{Network property} & \textbf{Sampling method} & \textbf{Embedding model} & \textbf{Optimization technique} \\
		\hline
		\multirow{7}[34]{*}{\tabincell{c}{Conventional \\ NRL}} & MVE \cite{DBLP:conf/cikm/QuTSR0017} & -, Adjacency matrix & \tabincell{c}{Multi-view first- \\ order proximities} & Edge sampling & Skip-gram model & Attention mechanism, negative sampling \\
		\cline{2-7}          & SDNE \cite{DBLP:conf/kdd/WangC016} & {Matrix-based processing, Adjacency matrix,  Laplacian matrix} & 1st, 2nd order proximity & {Fetch from adjacency matrix} & {Deep autoencoder, Laplacian Eigenmaps} & Deep belief network \\
		\cline{2-7}          & DNGR \cite{DBLP:conf/aaai/CaoLX16} & {Random surfing, PCO matrix} & High-order proximity & {Fetch from PPMI matrix} & {SDAE} & Negative sampling \\
		\cline{2-7}          & Struc2vec \cite{DBLP:conf/kdd/RibeiroSF17} & {Graph decomposition,  Context graph} & Node identity proximity & {Biased random walk} & Skip-gram model & {Hierarchical softmax} \\
		\cline{2-7}          & MRF \cite{DBLP:conf/aaai/JinYLHCFC19} & {Matrix-based processing,} Transition matrix & Community structure & Random pairwise sampling & Markov random field & Gaussian mixture model  \\
		\cline{2-7}          & splitter \cite{DBLP:conf/www/EpastoP19} & Graph decomposition, {Persona graph} & Overlapping community structure & Random walk & Skip-gram model & Hierarchical softmax \\
		\cline{2-7}          & vGraph \cite{DBLP:conf/nips/SunQHH019} & -, Adjacency matrix & Community structure  & Edge sampling & Probabilistic generative model & Negative sampling  \\
		\hline
		\multirow{2}[10]{*}{\tabincell{c}{Distributed \\ NRL}} & {GraphVite \cite{DBLP:conf/www/ZhuXTQ19}} & {-, Adjacency matrix} & {1st, 2nd order proximity} & {Parallel random walk} & {Edge probabilistic model} & {Parallel negative sampling} \\
		\cline{2-7}          & {PBG \cite{DBLP:sysml/abs-1903-12287}} & { Graph decomposition, Multi-bucket } & { 1st-order proximity } & { Distributed edge sampling } & { Edge probabilistic model } & { batched negative sampling } \\
		\hline
		\multirow{3}[18]{*}{Multi-NRL} & { DMNE \cite{DBLP:conf/www/NiCLCCX018}} & {-, Adjacency matrix, cross-network relationship matrix } & { High-order proximity under many-to-many mapping } & { Random walk } & { Deep autoencoder } & { Negative sampling } \\
		\cline{2-7}          & TransLink \cite{DBLP:conf/infocom/ZhouF19} & Meta-path extraction, {Meta-path,} node state, edge state & High-order proximity under translations & Fetch from adjacency matrix & Translating embedding & Negative sampling  \\
		\cline{2-7}          & CENALP \cite{DBLP:conf/ijcai/DuYZ19} &{-, Adjacency matrix} & High-order proximity & {Biased random walk}& Skip-gram model & Negative sampling \\
		\hline
		\multirow{5}[34]{*}{Dynamic NRL} & GraphSage \cite{DBLP:conf/nips/HamiltonYL17} &{Matrix-based processing,} Transition matrix, node state & 2nd-order proximity & Random walk & Spatial GCN & Negative sampling  \\
		\cline{2-7}          & Graph2Gauss \cite{DBLP:conf/iclr/BojchevskiG18} &{-, Adjacency matrix, node state} & High-order proximity & Node-anchored sampling & Deep autoencoder  & Negative sampling  \\
		\cline{2-7}          & SPINE \cite{DBLP:conf/ijcai/GuoXL19} &{Matrix-based processing, Rooted PageRank matrix, node state} & Node identity proximity & Random walk & Skip-gram model & {Biased positive sampling,} Negative sampling  \\
		\cline{2-7}          & { DynamicTriad \cite{DBLP:conf/aaai/ZhouYR0Z18}} & {-, Adjacency matrix } & { 1st-order proximity } & { Edge sampling } & { Edge probabilistic model } & { Negative sampling base on corruption} \\
		\cline{2-7}          & { DyRep \cite{DBLP:conf/iclr/TrivediFBZ19}} & {-, Adjacency matrix, node state, edge state } & { 2nd-order proximity } & { Random walk } & { GNN } & { Graph attention network, negative sampling } \\
		\hline
		\multirow{3}[8]{*}{\tabincell{c}{KRL}} & { RotatE \cite{DBLP:conf/iclr/SunDNT19} } & { -, Adjacency matrix } & { Distance proximity } & { Fetch from adjacency matrix} & { Translating embedding } & { Self-adversarial negative sampling } \\
		\cline{2-7}          & { SACN \cite{DBLP:conf/aaai/ShangTHBHZ19} } & { -, Adjacency matrix, node state, edge state } & { Distance proximity } & { Fetch from adjacency matrix } & { Weighted GCN } & { Negative sampling } \\		
		\hline
	\end{tabular}%
	\label{tab6}%
\end{table}%

\subsection{Discussions and Open Challenges}
The past half-decade has witness the rapid development of NRL, meanwhile NRL research also faces multiple challenges, among which we select some promising challenges with high attention to discuss, hoping to provide useful guidance for future study.

\subsubsection{Automated Learning}
We have shown that different NRL approaches tend to exhibit different performances for different scenarios. There is no single method that is the winner for all scenarios. When one method is considered to be better than another, it typically depends on which benchmark is used, including tasks and datasets available for evaluation. According to our framework, proper combinations of candidate methods, models, and techniques at different stages of NRL may have greater potential to boost the performance of NRL. For example, Struc2vec \cite{DBLP:conf/kdd/RibeiroSF17} preprocesses the input graph into a context graph, generates samples via a biased random walk method and learns representations by leveraging skip-gram model as well as negative sampling technique to reflect features w.r.t. node identities. It ranks the second-best model for node classification on Wikipedia \cite{struc2vecwiki}.
Most of existing NRL approaches follow the data-driven paradigm and is semi-automatic, i.e., given a benchmark with the specified dataset, we have to decide 1) how the data should be preprocessed? 2) what features should be preserved? 3) what models and techniques can be used? 4) how to optimize the chosen model? It is not a trivial task even for an expert to answer the above questions. Some recent efforts on meta-learning \cite{DBLP:journals/corr/abs-2004-05439} hold potential to partially address these questions. The learning-to-learn mode allows meta-learning being useful in multi-task scenarios where task-agnostic knowledge is learned from a family of tasks and used to improve learning of new tasks from the same family. 
AutoNE \cite{DBLP:conf/kdd/TuM0P019} presents an automatic hyperparameter optimization, the principle behind which is transferring the knowledge regarding optimized hyperparameters from multiple subnetworks to the original network.

Nevertheless, implementing a fully automated learning of network representation remains an open problem.
\subsubsection{Proximity vs. Distinguishability}
Observed from Table 6, we find that most of the current studies prefer to put feature proximity preserving as the guideline to learn node representations, so as to ensure structurally similar nodes having similar representations. As a result, application tasks like node classification and social recommendation benefit from this proximity-driven embedding. On the contrary, enhancing proximity reduces the distinguishability of nodes at the same time. {\color{black} Here the distinguishability refers to that an arbitrary node is significantly different from others in the representation space even though there is structural proximity between nodes. The enforced proximity imposes adverse impacts on some other application tasks.} For example, in order to accurately identify anchor users between different social networks, it is desirable to ensure that each node's representation should be explicitly separated from the representations of its proximities. Otherwise, it would be hard to discriminate the anchor nodes from their structurally similar neighborhoods. Generative Adversarial Networks (GANs) \cite{DBLP:conf/nips/GoodfellowPMXWOCB14} may provide a possible way to enhance distinguishability, e.g., dNAME \cite{DBLP:conf/infocom/0002WTZZL19} presents a GAN mechanism to learn the latent space of single network combined with a graph kernel based regularizer for discriminating anchor nodes from others. GAN's function depends on capturing true data distribution. How to make use of GANs to finely distinguish each node meanwhile preserving proximity is still {\color{black}open to exploring}. GCNs \cite{DBLP:journals/corr/BrunaZSL13} generate node representations by combining its own features with features aggregated from its neighborhoods. A well-designed convolution operation can benefit the trade-off between proximity and distinguishability. Nowadays, many real-world networks can be classified as scale-free networks such as airline networks and co-authorship networks. Scale-free networks follow power-law distributions and the vast majority of network nodes have few neighbors, on the contrary, a tiny fraction of network nodes have huge number of neighbors. {\color{black} Can we design a proper preprocessing method to deal with such extreme imbalance and to keep the structural features of different nodes at the same time? The receptive field for convolution largely depends on the adopted sampling method. How to design a sampling method to collaborate with preprocessing method and embedding models to handle this challenge still needs to be explored. }
\subsubsection{Interpretability}
Compared with handcrafted feature engineering, the superiority of NRL research has been empirically verified by both visualization and benchmarks, but less studies on NRL can give theoretically satisfactory answers to the following fundamental questions: 
1) what exactly latent features are learned from the network? 
2) what contributes to good performance on visualization and benchmarks? 
The connections between the underlying working mechanism and the performance results have not been well revealed yet. 
NetMF \cite{DBLP:conf/wsdm/QiuDMLWT18} makes a good attempt to create intrinsic connections between four NRL methods (e.g., DeepWalk \cite{DBLP:conf/kdd/PerozziAS14}, LINE \cite{DBLP:conf/www/TangQWZYM15}, PTE \cite{DBLP:conf/kdd/TangQM15} and node2vec \cite{DBLP:conf/kdd/GroverL16}) and graph Laplacian by unifying them into the matrix factorization model.
{\color{black} Gogoglou et al. \cite{DBLP:journals/corr/abs-1910-03081} try to explore the association of the embedding space with both external and internal node categorization. Abductive learning \cite{DBLP:conf/nips/DaiX0Z19} provides another 
beneficial attempt by unifying machine learning and logic programming, where the former learns to perceive primitive logic facts from data while the latter is able to exploit symbolic domain knowledge and correct the wrongly perceived facts for improving learning results including interpretability.}
In short, to answer the above questions and ensure that the learned representations {\color{black}truly} reflect the network information, more efforts need to be devoted to the interpretable learning in the future. 


\section{Conclusions}
We have presented a comprehensive overview of the state-of-the-art NRL techniques through a unifying reference framework. We describe and compare different NRL approaches in terms of the network data preprocessing methods, the network feature learning models and the node embedding and optimization techniques. We also discuss open challenges of NRL research, including automated learning, trade-off between proximity and distinguishability and interpretability. 
Our unified three phase reference framework offers a number of unique benefits: 
	(1) This framework presents a new perspective to reviewing the state of the art NRL methods and technical optimizations employed.
	(2) The unified framework promotes reviewing the state of the art NRL methods using three progressive and complementary phases. This new approach provides an in-depth understanding of network representation learning from three important perspectives: (i) raw input preprocessing, (ii) the node feature extraction task and different methods utilized for improving and optimizing node feature extraction qualities, and (iii) network representation learning models.    
	(3) This unified three phase framework by design serves dual purposes: (i) help beginner readers and practitioners to understand the end to end NRL workflow, and (ii) help researchers and graduate students who are interested in learning the state of the art NRL to gain a deeper understanding through reviewing each of the three phases in the network representation learning workflow.


\begin{acks}
The first author conducted this work during his one year visit at Georgia Institute of Technology under Jiangsu Overseas Visiting Scholar Program for University Prominent Young \& Middle-aged Teachers and Presidents. The authors from Soochow University acknowledge the support from National Natural Science Foundation of China under Grants 61972272, 62172291, 62072321, U1905211 and Open Project Program of the State Key Laboratory of Mathematical Engineering and Advanced Computing under Grant 2019A04. The authors from Georgia Institute of Technology acknowledge the partial support from the National Science Foundation under Grants 2038029, 1564097 and an IBM faculty award.
\end{acks}



\bibliographystyle{ACM-Reference-Format}
\bibliography{csur}

\end{document}